\documentclass[10pt]{elsarticle}

\usepackage{csquotes}

\usepackage[normalem]{ulem}
\usepackage{dirtytalk}
\usepackage{lineno,hyperref}
\usepackage{array, booktabs, makecell}
\usepackage[table]{xcolor} 
\usepackage{tcolorbox}
\usepackage{comment}
\usepackage{color,soul}
\usepackage{cleveref}
\usepackage{graphicx}
\usepackage{adjustbox}
\usepackage{caption}
\usepackage{subcaption}
\usepackage[title]{appendix}
\usepackage{bibentry}
\usepackage{tabularx}
    \newcolumntype{L}{>{\raggedright\arraybackslash}X}
\nobibliography*
\DeclareCaptionType{TextBox}
\usepackage{listings}
\usepackage{parcolumns}
\usepackage{dirtree}
\usepackage[T1]{fontenc}
\usepackage{longtable}
\usepackage{float}
\usepackage{bbding}
\usepackage{pifont}
\usepackage{wasysym}
\usepackage{amssymb}
\usepackage{rotating}
\usepackage{pstricks}
\usepackage{multirow}
\usepackage{pgfplots}
\pgfplotsset{compat=1.14}

\usepackage{tikz}
\usetikzlibrary{plotmarks}
\usepackage{verbatim}

\modulolinenumbers[5]

\journal{Journal of \LaTeX\ Templates}

\usepackage{tikz}
\usetikzlibrary{shapes.geometric, arrows}

\tikzstyle{startstop} = [rectangle, rounded corners, minimum width=3cm, minimum height=1cm,text centered, draw=black, fill=black!10]

\tikzstyle{io} = [trapezium, trapezium left angle=70, trapezium right angle=110, minimum width=3cm, minimum height=1cm, text centered,text width=2cm, draw=black]

\tikzstyle{process} = [rectangle, minimum width=3cm, minimum height=1cm, text centered, text width=3cm, draw=black]

\tikzstyle{decision} = [diamond, minimum width=3cm, minimum height=1cm, text centered,text width=2cm, draw=black]

\tikzstyle{arrow} = [thick,->,>=stealth]

\tikzstyle{database} = [cylinder, shape border rotate=90, draw=black,minimum height=2cm,minimum width=3cm, text centered,text width=0.6cm]

\usepackage{listings}

\newcommand{\lstbg}[3][0pt]{{\fboxsep#1\colorbox{#2}{\strut #3}}}
\lstdefinelanguage{diff}{
  basicstyle=\ttfamily\scriptsize,
  morecomment=[f][\lstbg{red!20}]-,
  morecomment=[f][\lstbg{green!20}]+,
  morecomment=[f][\textit]{@@},
}

\definecolor{javared}{rgb}{0.6,0,0} 
\definecolor{javagreen}{rgb}{0.25,0.5,0.35} 
\definecolor{javapurple}{rgb}{0.5,0,0.35} 
\definecolor{javadocblue}{rgb}{0.25,0.35,0.75} 
 
\usepackage{multirow} 
\usepackage{multicol}
\usepackage{varwidth} 

\newcolumntype{L}[1]{>{\raggedright\arraybackslash}m{#1}} 
\definecolor{Gray}{gray}{0.80} 

\newlength\q 

        {\begin{itemize}\setlength{\itemsep}{0pt}\setlength{\parsep}{0pt}
	\setlength{\topsep}{0pt}\setlength{\parskip}{0pt}}
	{\end{itemize}}
	{\begin{enumerate}\setlength{\itemsep}{0pt}\setlength{\parsep}{0pt}
	\setlength{\topsep}{0pt}\setlength{\parskip}{0pt}}
	{\end{enumerate}}
	{\begin{enumerate}\setlength{\itemsep}{0pt}\setlength{\parsep}{0pt}
	\setlength{\topsep}{0pt}\setlength{\parskip}{0pt}\setlength{\leftmargin}{0pt}}
	{\end{enumerate}}   

\begin{document}

\begin{frontmatter}

\title{
On Preserving the Behavior in Software Refactoring: A Systematic Mapping Study}


\author[RIT]{Eman Abdullah AlOmar\corref{mycorrespondingauthor}}
\cortext[mycorrespondingauthor]{Corresponding author}
\ead{eman.alomar@mail.rit.edu}

\author[RIT]{Mohamed Wiem Mkaouer}
\ead{mwmvse@rit.edu}

\author[RIT]{Christian Newman}
\ead{cdnvse@rit.edu}

\author[ETS]{Ali Ouni}
\ead{ali.ouni@etsmtl.ca}

\address[RIT]{Rochester Institute of Technology, Rochester, NY, USA}
\address[ETS]{ETS Montreal, University of Quebec, Montreal, QC, Canada}

\begin{abstract} 

\noindent\textbf{Context:} Refactoring is the art of modifying the design of a system without altering its behavior. The idea is to reorganize variables, classes and methods to facilitate their future adaptations and comprehension. As the concept of behavior preservation is fundamental for refactoring, several studies, using formal verification, language transformation and dynamic analysis, have been proposed to monitor the execution of refactoring operations and their impact on the program semantics. However, there is no existing study that examines the available behavior preservation strategies for each refactoring operation. 

\noindent\textbf{Objective:} This paper identifies behavior preservation approaches in the research literature.

\noindent\textbf{Method:} We conduct, in this paper, a systematic mapping study, to capture all existing behavior preservation approaches that we classify based on several criteria including their methodology, applicability, and their degree of automation. 

\noindent\textbf{Results:} The results indicate that several behavior preservation approaches have been proposed in the literature. The approaches vary between using formalisms and techniques, developing automatic refactoring safety tools, and performing a manual analysis of the source code.

\noindent\textbf{Conclusion:} Our taxonomy reveals that there exist some types of refactoring operations whose behavior preservation is under-researched. Our classification also indicates that several possible strategies can be combined to better detect any violation of the program semantics.
\end{abstract}

\begin{keyword}
refactoring \sep behavior preservation \sep systematic mapping study
\end{keyword}

\end{frontmatter}

\section{Introduction}
\label{sec:Introduction}

Software maintenance and evolution is an essential activity for any software system and the success of a system measures by its ability to maintain a high quality of design in the face of continuous changes. Because the change to the code base is inevitable, mechanisms must be employed in order to avoid causing deterioration to its integrity. One of the key mechanisms to cope with this challenge is refactoring. Refactoring is the process of optimizing the internal structure of the code without changing its external behavior. With the existence of many refactoring techniques, developers are still reluctant to rely on refactoring frameworks, and they prefer to refactor their code manually \cite{kim2014empirical,Silva:2016:WWR:2950290.2950305}. Surveys have revealed developer's lack of trust in automatic refactoring \cite{6112738}, due to the fear of breaking the code semantics and introducing bugs. Although refactoring, by definition, guarantees the safety and preservation of the refactored system's functionality, its adoption is still limited. One way to narrow the gap between refactoring and its adoption, is to highlight the existing effort in securing the execution of refactoring operations. However, little is known about how existing verification techniques allow such variety of changes, which vary from renaming methods and attributes, to extracting classes and merging packages, to be executed without altering the software's functionality. Thus, There is a lack of comprehensive studies to keep researchers and practitioners up-to-date with the status of research in preserving the behavior, evaluating the correctness of the transformation, and whether or not these approaches lead to a safe and trustworthy refactoring. 

While refactoring has been the focus on several SLRs, these studies have mainly focused on identifying refactoring opportunities, through the identification of code smells, as a detection step, and on recommending the appropriate refactoring operations, as a correction step. Our work is different from these papers since our SLM primarily focuses on collecting and summarizing all the behavior preservation techniques in all areas of software refactoring. It is not limited to design-based approaches; it also covers code-based behavior preservation approaches. To the best of our knowledge, no previous work has conducted a comprehensive SLM pertaining to behavior preservation techniques in software refactoring. 

The goal of this paper is to report an SLM that (1) identifies behavior preservation approaches in the research literature, and (2) identifies open issues in existing research. The outcomes of this SLM can serve as summarizing indexes and are expected to (1) assist researchers to identify related behavior preservation topics that are not well explored, and (2) guide practitioners to know the existing techniques for behavior preservation, which have an impact on refactoring decisions made in practice. 

To conduct this systematic mapping study, we followed established guidelines for SLR and SLM studies in SE \cite{Kitchenham07guidelinesfor,wohlin2014guidelines,petersen2008systematic}. We performed the review by defining the search string, the search academic article search engine, the selection criteria, and the research questions. We extracted data for 101 potentially relevant articles using the search academic article search engine. After careful screening of these articles, we identified 28 primary studies (PSs). We classified these PSs based on different perspectives, including the software artifacts and language paradigms, the refactoring operations, the behavior preservation approaches, and the evaluation methods considered. 
We identified several topics and challenges in need to be addressed in future research.

\section{Background \& Related Work}
\label{sec:Background_RelatedWork}
\subsection{Behavior preserving transformation}
Refactoring is a maintenance task in which the internal structure of the source code is improved  while the external behavior is preserved \cite{Fowler:1999:RID:311424}. The definition of behavior preservation, originally introduced by Opdyke \cite{Opdyke:1992:ROF:169783}, states that, \textit{“for the same set of input values, the resulting set of output values should be the same before and after the refactoring.”} Opdyke supports the notion of behavior preservation by specifying refactoring preconditions. An example of a refactoring precondition can be seen when considering \textit{Extract Class} refactoring in which naming conflicts must be avoided. Opdyke defined seven properties that must be checked before refactoring programs, which include: (1) unique superclass, (2) distinct class names, (3) distinct member names, (4) inherited member variables not redefined, (5) compatible signatures in member function redefinition, (6) type-safe assignment, and (7) semantically equivalent references and operations.  

Some refactoring techniques and formalisms to guarantee program preservation have been reported in a survey study  by Mens and Tourwe \cite{1265817}. They discussed the existing literature in terms of refactoring activities applied and their techniques, the application of refactoring to any type of software artifacts, refactoring tool support, and the impact of refactoring on the software process. In one of these several refactoring classification aspects, they discussed how the use of assertions (preconditions, postconditions, and invariants) and the use of graph transformation could help in guaranteeing behavior preservation. Therefore, in contrast to this SLM, the previous survey does not cover all of the approaches to guarantee behavior-preserving transformation. The survey considered only a few studies on behavior preservation because of its broader topic in the area of software refactoring and because it was performed a decade ago. 

\subsection{Other systematic literature reviews in refactoring}

\begin{table*}[ht!]
\begin{center}
\caption{Refactoring-related SLRs in Related Work.}
\label{Table:RelatedWorkSLR}
\begin{adjustbox}{width=1.0\textwidth,center}
\begin{tabular}{lllc}\hline
\toprule
\bfseries Study & \bfseries Year  & \bfseries Focus & \bfseries No. of PSs \\
\midrule
Zhang et al. \cite{Zhang:2011:CBS:1967084.1967086} & 2011 & Bad smells \& refactoring & 39 \\
Abebe and Yoo \cite{article2} & 2014 & Trends, opportunities \& challenges of software refactoring & 58 \\
Misbhauddin and Alshayeb \cite{Misbhauddin2015} & 2015 & UML model refactoring & 94 \\
AlDallal \cite{ALDALLAL2015231} & 2015 & Refactoring opportunities identification & 47 \\
Singh and Kaur \cite{SINGH2017} & 2017 & Refactoring opportunities identification & 238 \\
AlDallal and Abdin \cite{7833023} & 2017 & Impact of refactoring on quality & 76 \\
Mariani and Vergilio \cite{MARIANI201714} & 2017 & Search-based refactoring & 71 \\
Baqais and Alshayeb \cite{baqais2020automatic} & 2020 & Automatic refactoring & 41 \\
\bottomrule
\end{tabular}
\end{adjustbox}
\end{center}
\end{table*}

This work is a systematic mapping study in which we studied and summarized the primary studies (PSs) reporting the behavior preservation approach in the area of software refactoring. We did not find any SLM discussing the behavior preservation strategies. However, we reviewed a number of existing SLRs because of the similarities between those works and ours in terms of research setting. Table~\ref{Table:RelatedWorkSLR} summarizes the SLRs cited in this study. 

Zhang et al. \cite{Zhang:2011:CBS:1967084.1967086} conducted an SLR of 39 studies in the field of bad code smells. They discussed these studies based on the following: the goals of the studies, type of code smells addressed, the approaches to studying code smells, and identifying bad smells and refactoring opportunities. 

In a systematic review reported by Abebe and Yoo \cite{article2}, 58 studies were reviewed with the intention of revealing the trends, opportunities, and challenges of software refactoring. Their classification helped guide researchers to address the crucial issues in the field of software refactoring. 

Misbhauddin and Alshayeb \cite{Misbhauddin2015} performed an SLR in the area of refactoring UML models. They analyzed and classified 94 PSs based on several criteria: UML types of models, the formalisms used, and the effect of refactoring on model quality. In part of the research, they listed a few model behavior specification approaches. Our SLM is not limited to design-based approaches; it also covers code-based behavior preservation approaches. 

AlDallal \cite{ALDALLAL2015231} conducted an SLR of 47 PSs published on identifying refactoring opportunities in object-oriented code. AlDallal's review classified PSs based on the considered refactoring scenarios, the approaches to determine refactoring candidates, and the datasets used in the existing empirical studies. In a following SLR work by AlDallal and Abdin \cite{7833023}, they discussed 76 PSs and classified based on refactoring quality attributes of object-oriented code.

Singh and Kaur \cite{SINGH2017} performed an SLR as an extension of AlDallal's \cite{ALDALLAL2015231} SLR. In their review, they analyzed 238 research items in the field of code smell detection and its refactoring opportunities with the intention of addressing some research questions that were left open in AlDallal's SLR.

More recently, Baqais and Alshayeb \cite{baqais2020automatic} conducted a systematic literature review on automated software refactoring. In their review, they analyzed 41 studies that propose or develop different automatic refactoring approaches. 

In the area of search-based refactoring, Mariani and Vergilio \cite{MARIANI201714} systematically reviewed 71 studies and classified them based on the main elements of search-based refactoring, including artifacts used, encoding and algorithms used, search technique, metrics addressed, available tools, and conducted evaluation. Within the field of search-based refactoring, Mariani and Vergilio classified the selected PSs into five general categories related to behavior preservation methods. These categories involved: (1) Opdyke's function \cite{Opdyke:1992:ROF:169783}, (2) Cinn{\'e}ide's function \cite{cinneide2001automated}, (3) domain-specific, (4) no evidence of behavior preservation, and (5) do not mention the method. The current SLM does not overlap Mariani and Vergilio's SLR because this SLM entirely focuses on behavior preservation transformation in all areas of software refactoring, whereas Mariani and Vergilio's SLR mainly focused on search-based refactoring and discussed partially general behavior preservation methods. 


As shown in Table~\ref{Table:RelatedWorkSLR}, all the above-mentioned studies focus on either (1) detecting refactoring opportunities, through the optimization of structural metrics, or the identification of design and code defects, or (2) automating the generation and recommendation of the most optimal set of refactorings to improve the system's design while minimizing the refactoring effort, so that developers still can recognize their own design. Our work is different from these papers since our SLM primarily focuses on collecting and summarizing all of the behavior  preservation techniques in all areas of software refactoring. It is not limited to design-based approaches; it also covers code-based behavior preservation approaches. To the best of our knowledge, no previous work has conducted a comprehensive SLM pertaining behavior preservation techniques in software refactoring.

\section{Research Method}
\label{sec:Research Method}
\begin{figure}[htbp]
	\centering
    \includegraphics[width=\textwidth]{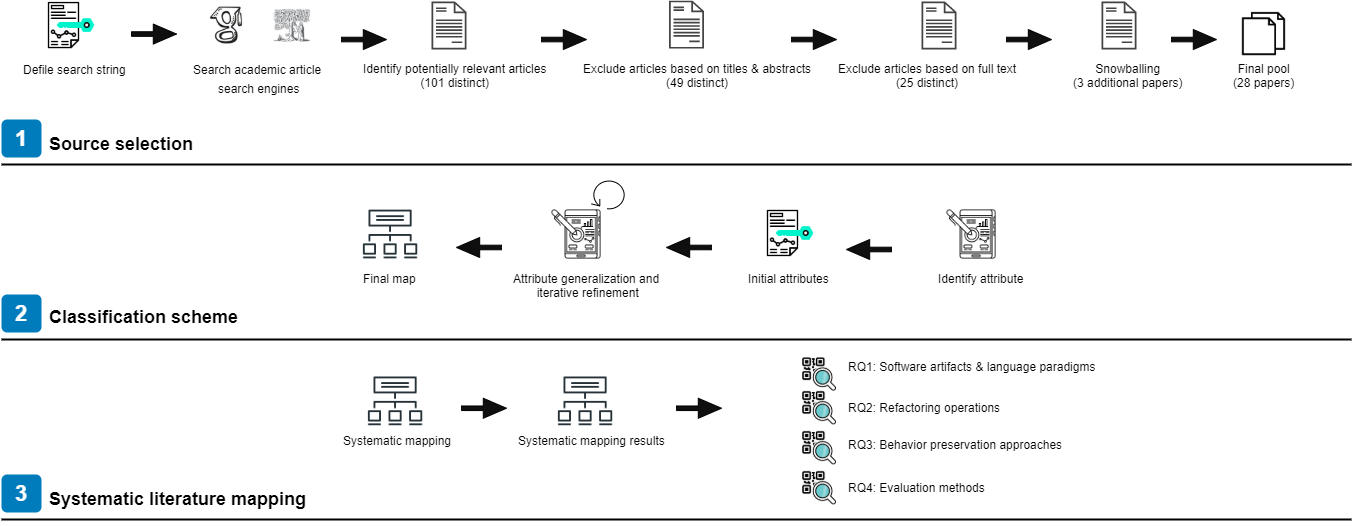}   
    \caption{Literature Search Process.}
    \label{fig:Literature_Search_Process}
\end{figure}

This SLM aggregates and summarizes the approaches in the field of testing behavior preservation in software refactoring. Based on the established guidelines \cite{Kitchenham07guidelinesfor,wohlin2014guidelines,petersen2008systematic}, we performed the SLM in three main phases: planning, conducting, and reporting the review. Creating a protocol is a major step when conducting an SLM \cite{Kitchenham07guidelinesfor}. This protocol contains the research questions, search strategy, study selection including inclusion and exclusion criteria, and data extraction and analysis to answer research questions. 

The core motivation behind carrying out this SLM is to: 
\begin{itemize}
\item Identify behavior preservation approaches in research literature.
\item Identify open issues in existing research.
\end{itemize} 

\subsection{Research questions}
Since little is known about the literature review of behavior preservation, this SLM serves as an exploration of this topic to extract existing techniques, currently being used, and their associated programming languages. The analysis of such wide variety of methods leads to develop a categorization and reveals areas of potential improvements. Therefore, we follow criteria defined in \cite{Kitchenham07guidelinesfor,wohlin2014guidelines,petersen2008systematic}when defining our research questions. The motivation behind each question is described below. 

\subsubsection{RQ1: What types of software artifacts and language paradigms were covered in the PSs to examine behavior preservation?}

The first research question explores the types of system levels and their language paradigms considered in the PSs, and to know what software artifacts are mostly used in the literature.

\subsubsection{RQ2: What refactoring types were considered in the PSs?}

Research question two identifies the refactoring operations that are tested and evaluated by behavior-preserving transformation approaches. 
This RQ serves as a popularity context to reveal the most and least popular refactoring types. Yet, the popularity in the context of behavior preservation is an indicator for refactoring complexity, as a code transformation, and thus, it potential proneness to errors. 

\subsubsection{RQ3: What approaches were considered by the PSs to test the behavior-preserving transformations in software refactoring?}

We pose this research question to study current approaches for testing behavior preservation of refactoring, and to get an overview of what different criteria are addressed by the existing methods. Accordingly, we collect information about refactoring techniques, automated analyses, and the manual analysis approach. Lastly, we check if the proposed approach is compared with existing methods, and study the pros and cons of the current approaches to suggest areas for improvement.


\subsubsection{RQ4: What evaluation methods were used in the PSs to assess the proposed behavior preservation approaches?}

We answer this research question by investigating how researchers evaluate and validate their proposed approaches in practice, when checking the reliability of the obtained conclusions. The answer to this question enumerates all evaluation methods that are found to be appropriate and most reliable when validating behavior preservation approaches.
\\

\subsection{Search strategy}
To find relevant studies, we performed an automatic search in Google Scholar and Scopus \footnote{www.scopus.com}. These search engines cover all main venues (e.g., IEEE, ACM, Springer). Our search string in these search engines was: 
\begin{TextBox}[h]
\centering
\fbox{\begin{minipage}{26em}
((behavior-preserving OR behavior preserving OR behavior preservation OR behaviour-preserving OR behaviour preserving OR behaviour preservation OR preserv* behavior OR preserv* behaviour) AND (formal OR method OR approach) AND (refactor* OR restructur*))
\end{minipage}}
\captionof{TextBox}{Search String.\label{txt:search_string}} 
\end{TextBox}

The strategy to construct search keywords was as follows:
\begin{itemize}
\item Derive the main terms from research questions and terms considered in the relevant papers.
\item Include alternative spellings for major terms.
\item Combine possible synonyms and spellings of the main terms using the Boolean OR operators, and then combine the main terms using the Boolean AND operators. 
\end{itemize}

These search keywords are applied to paper titles, abstracts, and keywords. To check the validity of the search string, we manually double check a few articles. Similar to \cite{GAROUSI2016195}, to restrict the search space when using Google Scholar to execute search string, we checked first several pages because we noticed that relevant studies appear in the first few pages.
 The process of determining the final list of PSs is depicted in Figure~\ref{fig:Literature_Search_Process}.

\subsection{Study selection}
To collect the PSs, we adapted the search process of \cite{7833023} and conducted a four phased process.
\subsubsection{Stage 1}                                   In this first stage of the paper selection process, given in Figure~\ref{fig:Literature_Search_Process}, we searched the academic article search engine for potentially related articles. Our criteria included applying our predefined search string against a publication's title, abstract, and keyword fields. Results from this search were not limited to specific venues. Searching the Google Scholar and Scopus resulted in a total of 101 literature publications. To reduce the possibility of including totally irrelevant articles, we performed the initial screening of the articles. Literature publications were then eliminated based on the defined inclusion and exclusion criteria to filter our irrelevant articles gathered in Stage 1. 

\textit{Inclusion criteria:}

The selected studies must satisfy all the following inclusion criteria:
\begin{itemize}
\item The article must be published in a peer-reviewed journal or conference before March 1, 2021.
\item The article must report an approach to testing behavior preservation and verify the correctness of refactorings. 
\end{itemize}

\textit{Exclusion criteria:}

Papers are excluded if satisfying any of the exclusion criteria, as follows:
\begin{itemize}
\item The study did not report an approach to test behavior-preserving transformations in software refactorings.
\item The study is a positioning paper, abstract, editorial, keynote, tutorial, or panel discussions. 
\item The study is not written in English
\end{itemize}

Regarding the second inclusion criteria, we only considered PSs that reported an approach to test the behavior preservation in refactoring, so we excluded any other articles that provided broad explanation about the concept of behavior preservation. Additionally, we excluded articles that were short because of their lack of comprehensiveness, e.g., \cite{article4}.

\subsubsection{Stage 2}
This stage involved an elimination of studies that were returned by the academic article search engine on the basis of the titles and abstracts of the potentially relevant articles. It is important to consider the abstracts in this stage because the titles of some articles could be misleading. The inclusion and exclusion rules were applied at this stage to all retrieved studies. This elimination process reduced our result set to  49 literature publications.

\subsubsection{Stage 3}
To obtain the relevant PSs, the complete literature publication was read and reviewed. Literature publications were eliminated based on the defined exclusion and inclusion rules. This process resulted in a total of 28 literature publications that were accepted for this study. 

\subsubsection{Stage 4}

To maximize the search coverage of all relevant papers, we conducted the snowballing technique \cite{wohlin2014guidelines} on papers already in the pool. It resulted in adding 3 additional papers, increasing the pool size to 28.

\subsection{Data extraction}
In order to determine the attribute(s) of the classification dimension, we screened the full texts of the PSs and identified the attribute(s) of that dimension. We used attribute(s) generalization and refinement to derive the final map, similar to \cite{GAROUSI2016195}. After the extraction of the classification dimension, we read the selected PSs in detail to answer the research questions. We then extracted the standard information from each paper, similar to \cite{KITCHENHAM20132049}, and included the additional attributes relevant to our study to the form. The data extraction form used is shown in Table~\ref{Table:Data_Extraction_Form}. 

Data stored in [F1] to [F11] are for documentation purposes, whereas data in [F12] to [F21] are for the purpose of data analysis. This form enables us to report the details needed for the PSs in this SLM.


\begin{table*}[h]
\begin{center}
\caption{Data Extraction Form.}
\label{Table:Data_Extraction_Form}
\begin{adjustbox}{width=0.9\textwidth,center}
\begin{tabular}{lll}\hline
\toprule
\bfseries No. & \bfseries Field & \bfseries Additional comments \\
\midrule
F1 & Primary study ID &  N/A \\
F2 & Author(s) &  N/A \\
F3 & Title &  N/A \\
F4 & Source &  N/A \\
F5 & Keyword &  N/A \\
F6 & Publication venue &  N/A \\
F7 & Type of publication & N/A \\
F8 & Date of publication &  N/A \\
F9 & Publication details for journal &  N/A \\
F10 & Citation count (Google Scholar) &  N/A \\
F11 & Page numbers &  N/A\\
F12 & Approach & Method used to ensure behavior preservation  \\
F13 & Approach Subcategory & A subcategory of each approach  \\
F14 & Strategy & A specific strategy used for that method to ensure behavior preservation   \\ 
F15 & Artifacts & System levels refactoring \\
F16 & Language Paradigm & A classification of software artifacts based on their features (if available) \\
F17 & Refactorings & List of refactoring scenarios  \\
F18 & Refactoring Classification & A classification for each refactoring operation \\
F19 & Evaluation Methods & A method used to validate and evaluate the proposed approach\\
F20 & Strength & A brief description of method's strengths \\ 
F21 & Limitation & A brief description of method's limitations \\
\bottomrule
\end{tabular}
\end{adjustbox}
\end{center}
\end{table*}

\section{Results}
\label{sec:Results}

\subsection{Overview of the PSs}
The research method discussed in Section~\ref{sec:Research Method} resulted in 28 relevant PSs listed in Appendix~\ref{sec:Appendix}. 
The main venues for these relevant PSs are presented in Table~\ref{Table:Publication_Sources}. The PSs were published in 18 different sources including journals and conferences. The list includes eleven journals and eleven conferences. The first relevant article discusses an approach of behavior preservation published in a journal in 1997, whereas the most recent one was published in 2018. The number of literature publication published in journals and conferences individually and combined are presented in Figure~\ref{fig:Distribution_PSs_by_year}.

Except for \cite{Tip:2003:RGU:949305.949308}, all the authors of the PSs are from academia. The authors of \cite{Tip:2003:RGU:949305.949308} are from industry. This indicates that most of the studies in this area were performed within an academic environment. 

Table~\ref{Table:Cites_by_Article.} shows an overview of the most-cited articles which indicate the degree of the most impactful PSs.

\begin{table*}[h]
\begin{center}
\caption{Publication Sources.}
\label{Table:Publication_Sources}
\begin{adjustbox}{width=1.0\textwidth,center}
\begin{tabular}{llll}\hline
\toprule
\bfseries Study & \bfseries Year & \bfseries Venue & \bfseries Source \\
\midrule
Roberts et al. \cite{roberts1997refactoring} & 1997 & Journal & Theory and Practice of Object Systems (TAPOS)\\
Mens et al. \cite{Mens2003FormalisingRW} & 2003 & Journal & Journal of Software Maintenance and Evolution (SME) \\
Tip et al. \cite{Tip:2003:RGU:949305.949308} & 2003 & Conference & Conference on Object-Oriented Programming, Systems, Languages, and Applications (OOPSLA) \\
Garrido and Meseguer \cite{4026866} & 2006 & Conference & International Workshop on Source Code Analysis and Manipulation (SCAM) \\
Straeten et al. \cite{VanDerStraeten2007} & 2007 & Journal & Software and System Modeling (SSM) \\
Massoni et al. \cite{Massoni:2008:FMP:1792838.1792873} & 2008 & Conference & Fundamental Approaches to Software Engineering (FASE) \\
Soares et al. \cite{article} & 2009 & Conference & Brazilian Symposium on Software Engineering (SBES) \\
Ubayashi et al.  \cite{ubayashi2008contract} & 2008 & Conference & International Conference on Software Testing, Verification, and Validation (ICST)\\
Sch{\"a}fer et al. \cite{schafer2008sound} & 2008 & Conference & Conference on Object-Oriented Programming, Systems, Languages, and Applications (OOPSLA) \\
Soares et al. \cite{soares2009generating} & 2009 & Conference & Brazilian Symposium on Programming Languages (SBLP) \\
Tsantalis and Chatzigeorgiou \cite{4752842} & 2009 & Journal & IEEE Transactions on Software Engineering (TSE) \\
Sch{\"a}fer and Moor \cite{schafer2010specifying} & 2010 & Conference & Conference on Object-Oriented Programming, Systems, Languages, and Applications (OOPSLA) \\
Soares et al. \cite{5440166} & 2010 & Journal & IEEE Software \\
Tsantalis and Chatzigeorgiou \cite{tsantalis2010identification} & 2010 & Journal  & Journal of Systems and Software (JSS)\\
Tip et al. \cite{tip2011refactoring} & 2011 & Journal & Transactions on Programming Languages and Systems (TOPLAS) \\
Soares et al. \cite{soares2011making} & 2011 & Conference & Brazilian Symposium on Programming Languages (SBLP)\\
Overbey and Johnson \cite{6100067} & 2011 & Conference & International Conference on Automated Software Engineering (ASE) \\
Soares et al. \cite{6080784} & 2011 & Conference & International Conference on Software Maintenance and Evolution (ICSME) \\
Soares et al. \cite{SOARES20131006} & 2013 & Journal & Journal of Systems and Software (JSS) \\
Jonge and Visser \cite{de2012language} & 2012 & Conference & Workshop on Language Description (WLD) \\
Noguera et al. \cite{noguera2012refactoring} & 2012 & Conference & International Conference on Software Maintenance and Evolution (ICSME) \\
Thies and Bodden \cite{thies2012refaflex} & 2012 & Conference & International Symposium on Software Testing and Analysis (ISSTA) \\
Mongiovi et al. \cite{Mongiovi:2014:MRS:2664678.2664813} & 2014 & Journal & Science of Computer Programming (SCP) \\
Najaf et al. \cite{najafi2016set} & 2016 & Journal & Computing and Informatics (CI)  \\
Horpácsi et al. \cite{Horpcsi2017TrustworthyRV} & 2017 & Conference & Verification and Program Transformation (VPT) \\
Mongiovi et al. \cite{7898404} & 2017 & Journal & IEEE Transactions on Software Engineering (TSE) \\ 
Chen et al. \cite{chen2018improving} & 2018 & Journal & Information and Software Technology (IST) \\
Insa et al. \cite{insa2018behaviour} & 2018 & Journal  & Scientific Programming (SP) \\

\bottomrule
\end{tabular}
\end{adjustbox}
\end{center}
\end{table*}

\begin{figure}[h!]  
  \begin{tikzpicture}
   \begin{scope}[scale=0.76]
     \begin{axis}[
       width=1.1\columnwidth,
       height=0.7\columnwidth,
       font=\sffamily,
       axis x line=bottom,
       xmin=1997,
       xmax=2018,
       xlabel={Year},
       xlabel near ticks,
       xticklabel style={/pgf/number format/1000 sep=},
       axis y line=left,
       ymin=0,
       ymax=10,
       ylabel={No. of PSs},
       ylabel near ticks
     ]
      \addplot[color=blue,mark=triangle*] coordinates {
         (1997,1)
         (1998,0)
         (1999,0)
         (2000,0)
         (2001,0)
         (2002,0)
         (2003,1)
         (2004,0)
         (2005,0)
         (2006,1)
         (2007,0)
         (2008,3)
         (2009,2)
         (2010,1)
         (2011,3)
         (2012,3)
         (2013,0)
         (2014,0)
         (2015,0)
         (2016,0)
         (2017,1)
         (2018,0)
       };
      \addplot[color=red,mark=square*,mark options={fill=white}] coordinates {
         (1997,0)
         (1998,0)
         (1999,0)
         (2000,0)
         (2001,0)
         (2002,0)
         (2003,1)
         (2004,0)
         (2005,0)
         (2006,0)
         (2007,1)
         (2008,0)
         (2009,1)
         (2010,2)
         (2011,1)
         (2012,0)
         (2013,1)
         (2014,1)
         (2015,0)
         (2016,1)
         (2017,1)
          (2018,2)
       };
       \addplot[color=black,mark=*] coordinates {
         (1997,1)
         (1998,0)
         (1999,0)
         (2000,0)
         (2001,0)
         (2002,0)
         (2003,2)
         (2004,0)
         (2005,0)
         (2006,1)
         (2007,1)
         (2008,3)
         (2009,3)
         (2010,3)
         (2011,4)
         (2012,3)
         (2013,1)
         (2014,1)
         (2015,0)
         (2016,1)
         (2017,2)
         (2018,2)
       };
       
        legend style={
    	at={(0.5,-0.05)},
        anchor=north
        },
        \legend{\# of conference PSs,\# of journal PSs,\# of all PSs}
     \end{axis}
     \end{scope}
   \end{tikzpicture}
   \caption{Distribution of Primary Studies by Year.} 
\label{fig:Distribution_PSs_by_year}
 \end{figure}
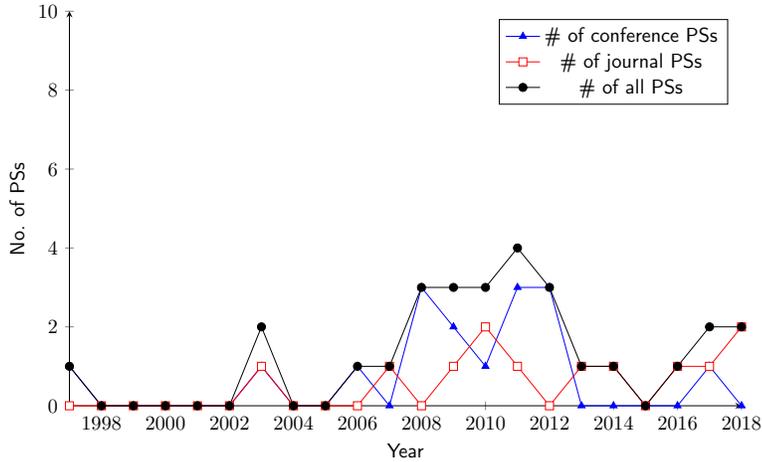
\begin{table}[h]
\begin{center}
\caption{Citation Count (Obtained from Google Scholar).}
\label{Table:Cites_by_Article.}
\begin{adjustbox}{width=0.5\textwidth,center}
\begin{tabular}{llll}
\toprule
\bfseries  Study  & \bfseries  Year & \bfseries Source & \bfseries Count \\
\midrule
Roberts et al. \cite{roberts1997refactoring} & 1997 & TAPOS & 550 \\
Tsantalis \& Chatzigeorgiou \cite{4752842}  & 2009 & TSE & 316 \\ 
Mens et al. \cite{Mens2003FormalisingRW} & 2003 & SME & 206\\ 
Tip et al. \cite{Tip:2003:RGU:949305.949308} & 2003 & OOPSLA & 181 \\ 
Soares et al. \cite{5440166} & 2010 &IEEE Software & 122 \\
Sch{\"a}fer \& Moor \cite{schafer2010specifying}& 2010 & OOPSLA & 104 \\
Sch{\"a}fer et al. \cite{schafer2008sound} & 2008 & OOPSLA & 103 \\
Tip et al. \cite{tip2011refactoring} & 2011 & Journal  & 74 \\
Straeten et al. \cite{VanDerStraeten2007} & 2007 & SSM & 64 \\
Tsantalis \& Chatzigeorgiou \cite{tsantalis2010identification} & 2010 & JSS & 59 \\
Garrido \& Meseguer \cite{4026866} & 2006 & SCAM & 49 \\
Soares et al. \cite{SOARES20131006} & 2013 & JSS & 37 \\
Overbey \& Johnson \cite{6100067} & 2011 & ASE & 33 \\
Mongiovi et al. \cite{Mongiovi:2014:MRS:2664678.2664813} & 2014 & SCP & 31 \\
Soares et al. \cite{6080784} & 2011 & ICSME & 29 \\
Thies and Bodden \cite{thies2012refaflex} &  2012 &  ISSTA &  23 \\
Massoni et al. \cite{Massoni:2008:FMP:1792838.1792873} & 2008 & FASE & 21 \\
\bottomrule
\end{tabular}
\end{adjustbox}
\end{center}
\end{table}

\subsection{RQ1: What types of software artifacts and language paradigms were covered in the PSs to examine behavior preservation?}
\label{Result:RQ1}

Table~\ref{Table:Software_Artifacts_Language_Paradigms.} presents the types of software artifacts, different language paradigms and programming and modeling languages used in the PSs. Refactoring is applied to only two kinds of artifacts in the literature: code and model. Code refactoring targets to apply refactoring techniques at the source code level. Model refactoring aims to apply refactorings at model level as opposed to the source code. 
 Most (82.75\%) of the PSs were about refactored source code, and a few (17.24\%) concerning refactored design models. Articles optimizing code are primarily focused on Java programming language. Few articles, however, used C++, Smalltalk, AspectJ, Fortran, PHP, BC, Erlang, Stratego, Mobl, and XML to test behavior preservation. For model refactoring, research deals with Alloy specification language or UML models. As can be seen from the table, most of the papers consider refactoring source code, focusing primarily on the Java language. Model refactoring is being used by few articles. Moreover, one of the articles \cite{Mens2003FormalisingRW} does not explicitly mention what types of artifacts were refactored. By analyzing the PS \cite{Mens2003FormalisingRW}, it is possible to guess that it is applicable to either code or models since the behavior preservation approach described is about graph transformation. 

The focus on Java language might be because of the popularity of Java, and refactoring examples in Fowler's book are written in Java. Researchers are encouraged to focus on different languages and apply more refactoring to design models when testing behavior preservation. 

\begin{table*}[h]
\begin{center}
\caption{Software Artifacts and its Language Paradigms.}
\label{Table:Software_Artifacts_Language_Paradigms.}
\begin{adjustbox}{width=0.9\textwidth,center}
\begin{tabular}{p{3cm}  p{4cm} p{4cm} p{6cm} }\hline
\toprule
\multicolumn{1}{l}{\bfseries {Software Artifact}}  & \bfseries Language Paradigm  & \bfseries Language & \bfseries  PSs  \\
\midrule
\multirow{4}{*}{Code} & Class-based OO & Java & \cite{Mongiovi:2014:MRS:2664678.2664813}\cite{article}\cite{soares2009generating}\cite{5440166}\cite{4752842}\cite{4026866}\cite{Tip:2003:RGU:949305.949308}\cite{6080784}\cite{7898404}\cite{SOARES20131006} \cite{chen2018improving}  \cite{thies2012refaflex}  \cite{tsantalis2010identification} \cite{de2012language} \cite{noguera2012refactoring} \cite{tip2011refactoring} \cite{schafer2010specifying} \cite{schafer2008sound} \cite{tip2011refactoring} \\ 
& & C++ & \cite{4752842}\\
& & Smalltalk & \cite{roberts1997refactoring}\\
& Aspect-oriented & AspectJ & \cite{Mongiovi:2014:MRS:2664678.2664813} \cite{soares2011making} \cite{ubayashi2008contract} \cite{noguera2012refactoring} \cite{schafer2008sound} \\ 
& Imperative & Fortran, PHP, BC & \cite{6100067} \\ 
& Functional & Erlang & \cite{Horpcsi2017TrustworthyRV} \cite{insa2018behaviour} \\ 
&  Domain-specific &  Stratego, Mobl & \cite{de2012language} \\
& Markup & XML & \cite{noguera2012refactoring} \\
& & &  \\
\multirow{2}{*}{Model} & Structural \& Behavioral & UML & \cite{VanDerStraeten2007} \cite{najafi2016set} \\ 
& Structural formal & Alloy & \cite{Massoni:2008:FMP:1792838.1792873}  \\
\bottomrule
\end{tabular}
\end{adjustbox}
\end{center}
\end{table*}

\begin{tcolorbox}
\textit{Summary.} Refactoring studies cover mainly two levels of artifacts: Source code and model. Source code artifacts are the main focus of refactoring literature. Java is the most popular programming language in these studies.
\end{tcolorbox}

\subsection{RQ2: What refactoring types were considered in the PSs?} 
\label{Result:RQ2}
As shown in Table~\ref{Table:Refactorings_Identified_by_PSs} and Table~\ref{Table:Refactorings_Identified_by_PSs_2}, the literature publications addressed 150 distinct refactoring operations. In this SLM, we classify refactoring operations considered in the PSs into three categories: Fowler's catalog, Model refactorings, and Language-specific refactorings. Refactorings proposed by Fowler fall into the first category (23 PSs), refactoring scenarios applied in design model fall into the second category (3 PSs), and the third category is assigned to refactorings that were applied by specific programming languages involved in the PS (14 PSs). It is important to note that some studies used refactoring operations that belong to two categories. 43 out of 150  refactoring activities were cataloged by Fowler \cite{Fowler:1999:RID:311424} and serve different purposes:  composing methods, organizing data, simplifying conditional expressions and method calls, dealing with generalization, and moving features between objects. The other refactorings are either model refactorings or language-specific associated with model or source code artifacts.  

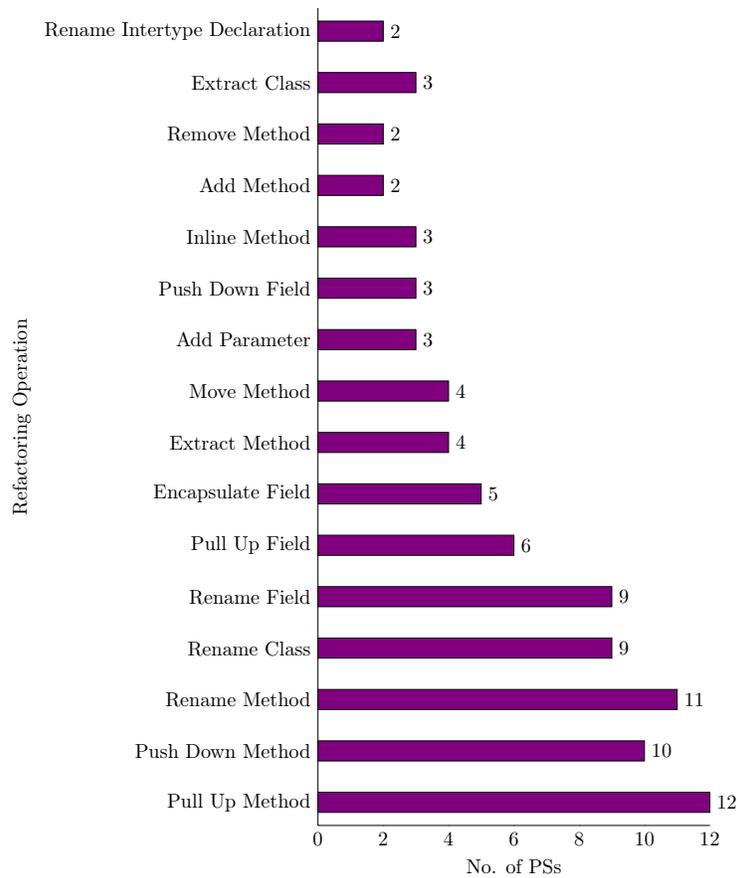
\begin{figure}[h]  
\centering 
\begin{tikzpicture}
\begin{scope}[scale=0.76]
  \begin{axis}[
    xbar,
    y axis line style = { opacity = 1 },
    axis x line*=bottom,
    axis y line*=left,
    xmin=0, xmax=12,
    tickwidth         = 0pt,
    tick align=inside,
    y=0.9cm,
    enlarge y limits  = 0.03,
    enlarge x limits  = 0,
    symbolic y coords = { 
        Pull Up Method,
        Push Down Method,
        Rename Method,
        Rename Class,
        Rename Field,
        Pull Up Field,
        Encapsulate Field,
        Extract Method,
        Move Method,
        Add Parameter, 
        Push Down Field,
        Inline Method,
        Add Method,
        Remove Method, 
        Extract Class,
        Rename Intertype Declaration,
   	},
   ylabel={Refactoring Operation},
   xlabel={No. of PSs},
   legend image code/.code={%
   \draw[#1, draw=none,/tikz/.cd,yshift=-0.25em]
        (0cm,1pt) rectangle (6pt,7pt);},
   legend style={
    	at={(0.2,-0.1)},
        anchor=north,
        legend columns=-1,
        /tikz/every even column/.append style={column sep=0.5cm}
        },
    nodes near coords
  ]
   \addplot[fill=violet] coordinates {
        (12,Pull Up Method)
        (11,Rename Method)
         (10,Push Down Method)
        (9,Rename Class)
        (9,Rename Field)
        (6,Pull Up Field)
        (5,Encapsulate Field)
        (4,Extract Method)
        (4,Move Method)
        (3,Add Parameter)  
        (3,Push Down Field)
        (3,Inline Method)
        (3,Extract Class)
        (2,Remove Method)
        (2,Add Method)
        (2,Rename Intertype Declaration)
   }; 
 
  \end{axis}
\end{scope}
\end{tikzpicture}
\caption{Distribution of Most Used Refactoring Operations.}
\label{Chart:RefactoringsOccurrence}
 \end{figure}

As can be observed in Figure~\ref{Chart:RefactoringsOccurrence}, some of the refactoring scenarios were studied more frequently than others. The TOP 3 most studied refactoring types are \textit{Pull Up Method}, \textit{Rename Method}, and \textit{Push Down Method}. 

Interestingly, while it is expected that PSs opt for popular refactorings, to guarantee their correctness, recent studies that have been mining refactorings \cite{tsantalis2018accurate,peruma2020contextualizing,alomar2021we,tsantalis2020refactoringminer} have shown that \textit{Pull Up Method}, and \textit{Push Down Method} are among the least used refactorings in practice.
We observe that the behavior can be preserved under less or very restrictive conditions depending on the nature of refactoring types. For example, when a class member (method or field) is moved up or down an inheritance hierarchy, or when it is required after the refactoring to have all references to the same variables and methods defined in the same class as before the refactoring as it seems these refactoring operations are the ones that most likely to introduce behavior changes. \textit{Pull Up Method}, and \textit{Push Down Method} are defined as the intention of moving identical methods or attributes, spread in subclasses, up into a superclass, or vice versa, respectively. These refactorings seem to be attractive for researchers to analyze, saying they can be highly useful when removing duplicate code, extracting reusable components, or implementing design patterns. Yet, since they impact several interconnected classes through hierarchies, it is critical to guarantee the refactoring execution correctness. However, developers are found to be rarely performing these types of refactorings through the IDE, instead, it is most likely that they manually move whatever member across hierarchies, and manually fix any unexpected errors that it may cause.

\begin{tcolorbox}
\textit{Summary.} A variety of refactoring operations have been used in the literature. These refactorings can be classified into three categories: Fowler’s catalog, model refactorings, and language-specific refactorings. When testing the proposed behavior preservation approaches by PSs, we observe that some refactoring types such as \textit{Pull Up Method}, \textit{Rename Method}, and \textit{Push Down Method} were studied more frequently than others. The high interest in these refactoring operations in primary studies may indicate their importance in preserving the behavior. 
\end{tcolorbox}

\begin{table*}[!htbp]
\begin{center}
\caption{Refactorings Identified by Primary Studies and their Classification Schema.}
\label{Table:Refactorings_Identified_by_PSs}
\begin{adjustbox}{width=1.0\textwidth,center}
\begin{tabular}{lccc}
\toprule
\multirow{2}{*}[-3pt]{\textbf{Refactorings}}   & \multicolumn{3}{c}{\textbf{Classification}} \\ 
        \cmidrule{2-4} 
      &   \textbf{Fowler's catalog} & \textbf{Model refactorings} & \textbf{Language-specific refactorings} \\ 
         \midrule
    Encapsulate Field  & \checkmark &  &  \\ 

    Pull Up Method & \checkmark &   &  \\ 

    Push Down Method & \checkmark &  &   \\ 
    Pull Up Field & \checkmark &  &   \\
    Rename Temporary &  &  &  \checkmark \\
    Move States into Orthogonal Composite State &  & \checkmark &  \\
    Flatten States &  & \checkmark &   \\
    Add Subclass  &  & \checkmark &  \\
    Introduce Signature &  & \checkmark &  \\
    Introduce Generalization &  & \checkmark &   \\
    Introduce Subsignature &  & \checkmark &  \\
    Introduce Relation  &  & \checkmark &  \\
    Remove Optional Relation &  & \checkmark &  \\
    Remove Scalar Relation &  & \checkmark &  \\
    Split Relation &  & \checkmark &  \\
    Rename Class &  &  &  \checkmark \\
    Rename Field &  &  &  \checkmark \\
    Rename Local Variable &  &  & \checkmark  \\
    Rename Method  & \checkmark &  &  \\
    Extract Method & \checkmark &  &   \\
    Extract Class & \checkmark &  &  \\
    Move Class &  &  &  \checkmark \\
    Change Method Signature &  &  &  \checkmark \\
    Move Method & \checkmark &  &  \\
    Rename &  &  &  \checkmark \\
    Move &  &  &  \checkmark \\ 
    Introduce USE &  &  &  \checkmark \\
    Change Function Signature &  &  & \checkmark \\
    Introduce Implicit None &  &  &  \checkmark \\
    Add Empty Subprogram  &  &  & \checkmark \\
    Safe Delete &  &  & \checkmark \\
    Copy Up Method &  &  & \checkmark \\
    Extract Local Variable &  &  &  \checkmark \\
    Add Local Variable &  &  & \checkmark \\
    Introduce Block &  &  &  \checkmark \\
    Insert Assignment &  &  & \checkmark \\
    Move Expression &  &  & \checkmark \\
    Extract Function &  &  &  \checkmark \\
    Add Empty Function &  &  &  \checkmark \\
    Populate Function  &  &  & \checkmark \\
    Replace Expression &  &  &  \checkmark \\
    Push Down Field & \checkmark &  &  \\
    Rename Type &  &  &  \checkmark \\
    Replace Code with Method Call &  &  & \checkmark \\
    Move Operation to Listener &  &  & \checkmark \\
    Remove Unused Variable &  &  & \checkmark \\
    Change Instance Access to Static &  &  & \checkmark \\
    Remove Immutable Object Copy &  &  & \checkmark \\
    Replace Direct Access with Getter &  &  & \checkmark \\
    Replace Instance with isInstance &  &  & \checkmark \\
    Remove Parameter & \checkmark &  &  \\
    Replace Field with Method & \checkmark &  &   \\
        Decrease Method Visibility &  &  &  \checkmark \\
    Replace Direct Access with Setter &  &  & \checkmark \\
    Inline Temp & \checkmark &  &  \\
    Consolidate Duplicate Code Fragment & \checkmark &  &  \\
    Rename Constant &  &  & \checkmark \\
    Rename Local Variable &  &  & \checkmark \\
    Replace Generic Cast with classCast &  &  & \checkmark \\
    Replace Generic Cast with isInstance  &  &  & \checkmark \\
    Replace Method with Method Object & \checkmark &  &  \\
    Change Statement Order &  &  & \checkmark \\
     Swap Access Method &  &  & \checkmark \\
    Remove Duplicate Assignment &  &  & \checkmark \\
    Consolidate Conditional Expression & \checkmark &  &   \\
    Introduce Explaining Variable & \checkmark &  &   \\
       Remove Assignment to Parameters & \checkmark &  &  \\
    Increase Method Visibility &  &  &  \checkmark \\
    Replace if with Switch &  &  & \checkmark \\
    Replace Equivalent Method Call &  &  & \checkmark \\
     Introduce Null Object & \checkmark &  &  \\
    Replace Magic Number with Constant & \checkmark &  &  \\
     Wrap (Change) Expression &  &  &  \checkmark \\
        \bottomrule
\end{tabular} 
\end{adjustbox}
\end{center}
\end{table*}

\begin{table*}[!htbp]
\begin{center}
\caption{Refactorings Identified by Primary Studies and their Classification Schema (Cont'd).}
\label{Table:Refactorings_Identified_by_PSs_2}
\begin{adjustbox}{width=1.0\textwidth,center}
\begin{tabular}{lccc}
\toprule
\multirow{2}{*}[-3pt]{\textbf{Refactorings}}   & \multicolumn{3}{c}{\textbf{Classification}} \\ 
        \cmidrule{2-4} 
      &   \textbf{Fowler's catalog} & \textbf{Model refactorings} & \textbf{Language-specific refactorings} \\ 
         \midrule
    Extract to Function &  &  & \checkmark \\
    Extract to Variable &  &  &  \checkmark \\
    Outer Variable &  &  & \checkmark \\
    Variable to Function Parameter &  &  & \checkmark \\
    Rename Function &  &  & \checkmark \\
    Add Method &  &  & \checkmark \\
    Remove Method &  &  & \checkmark \\
    Change Method Body &  &  & \checkmark \\
    Change Method Modifier &  &  & \checkmark \\
    Add Field &  &  & \checkmark \\
    Remove Field &  &  & \checkmark \\
    Change Field Modifier &  &  & \checkmark \\
    Change Field Initializer &  &  & \checkmark \\
    Change Static Field Initializer &  &  & \checkmark \\
    Rename Intertype Declaration &  &  & \checkmark \\
    Inline Method & \checkmark &  &   \\
    Extract Exception Handler &  &  & \checkmark \\
    Infer Generic Type &  &  & \checkmark \\
    Replace Deprecated Code &  &  & \checkmark \\
    Extract Interface & \checkmark &  &  \\
    Extract Subclass & \checkmark &  &  \\
    Generalize Type &  &  & \checkmark  \\
    Add Variable & & & \checkmark  \\
    Create Accessors for a Variable & & & \checkmark \\
    Change all Variable refs to Accessors Calls & & & \checkmark \\
    Remove Class & & & \checkmark \\
    Move Method across Object Boundry & & & \checkmark \\
    Extract Code as Method & & & \checkmark \\
Change Abstract Class to Interface & & & \checkmark \\
Extract Feature into Aspect & & & \checkmark \\
Extract Fragment into Advice & & & \checkmark \\
Extract Inner Class to Standalone & & & \checkmark \\
Inline Class within Aspect & & & \checkmark \\
Inline Interface within Aspect & & & \checkmark \\
Move Field from Class to Inter-type & & & \checkmark \\
Move Method from Class to Inter-type & & & \checkmark \\
Replace Implements with Declare Parents & & & \checkmark \\
Split Abstract Class into Aspect and Interface & & & \checkmark \\
Extend Marker Interface with Signature & & & \checkmark \\
Generalize Target Type with Marker Interface & & & \checkmark \\ 
Introduce Aspect Protection  & & & \checkmark \\
Replace Inter-type Field with Aspect Map  & & & \checkmark \\
Inter-type Method with AspectMethod & & & \checkmark \\
Tidy Up Internal Aspect Structure & & & \checkmark \\
Extract Superaspect & & & \checkmark \\
Pull Up Advice & & & \checkmark \\
Pull Up Declare Parents & & & \checkmark \\
Pull Up Inter-type Declaration & & & \checkmark \\
Pull Up Marker Interface & & & \checkmark \\
Pull Up Pointcut & & & \checkmark \\
Push Down Advice & & & \checkmark \\
Push Down Declare Parents & & & \checkmark \\
Down Inter-type Declaration & & & \checkmark \\
Push Down Marker Interface & & & \checkmark \\
Push Down Pointcut & & & \checkmark \\
Conditional with Polymorphism & \checkmark & &  \\
Rename Package  & \checkmark & &  \\
Move Type  & \checkmark & &  \\
Extract Superclass & \checkmark & &  \\
Add Parameter & \checkmark & &  \\
Extract \& Move Method & \checkmark & &  \\
Extract \& Pull Up Method & \checkmark & &  \\
Move \& Rename Method & \checkmark & &  \\
Move Member Type To Toplevel & \checkmark \\
Move Member & \checkmark \\
Move Inner To Toplevel & \checkmark \\
Convert Anonymous To Nested & \checkmark \\
Move Instance Method & \checkmark \\
Extract Constant & \checkmark \\
Extract Temp & \checkmark \\
Inline Constant & \checkmark \\
Introduce Factory & \checkmark \\
Introduce Indirection & \checkmark \\
Introduce Parameter & \checkmark \\
Introduce Parameter Object & \checkmark \\ 
Promote Temp To Field & \checkmark \\


\bottomrule
\end{tabular} 
\end{adjustbox}
\end{center}
\end{table*}

\subsection{RQ3: What approaches were considered by the PSs to test the behavior-preserving transformations in software refactoring?} 
\label{Result:RQ3}
As discussed in RQ1 and RQ2, refactoring is not restricted to software code, but it also applies to model. Concerning refactoring types used to preserve the behavior, PSs used a variety of refactoring operations. However, as seen from Figure~\ref{Chart:RefactoringsOccurrence}, a number of refactoring operations receive considerable attention due to the fact that these are most likely introduce behavioral changes. Considering the types of software artifacts and refactoring operations used in the PSs, we report, in this section,  several approaches for testing behavior preservation of refactoring. 

\begin{figure*}[ht]
	\centering
    \includegraphics[width=\textwidth]{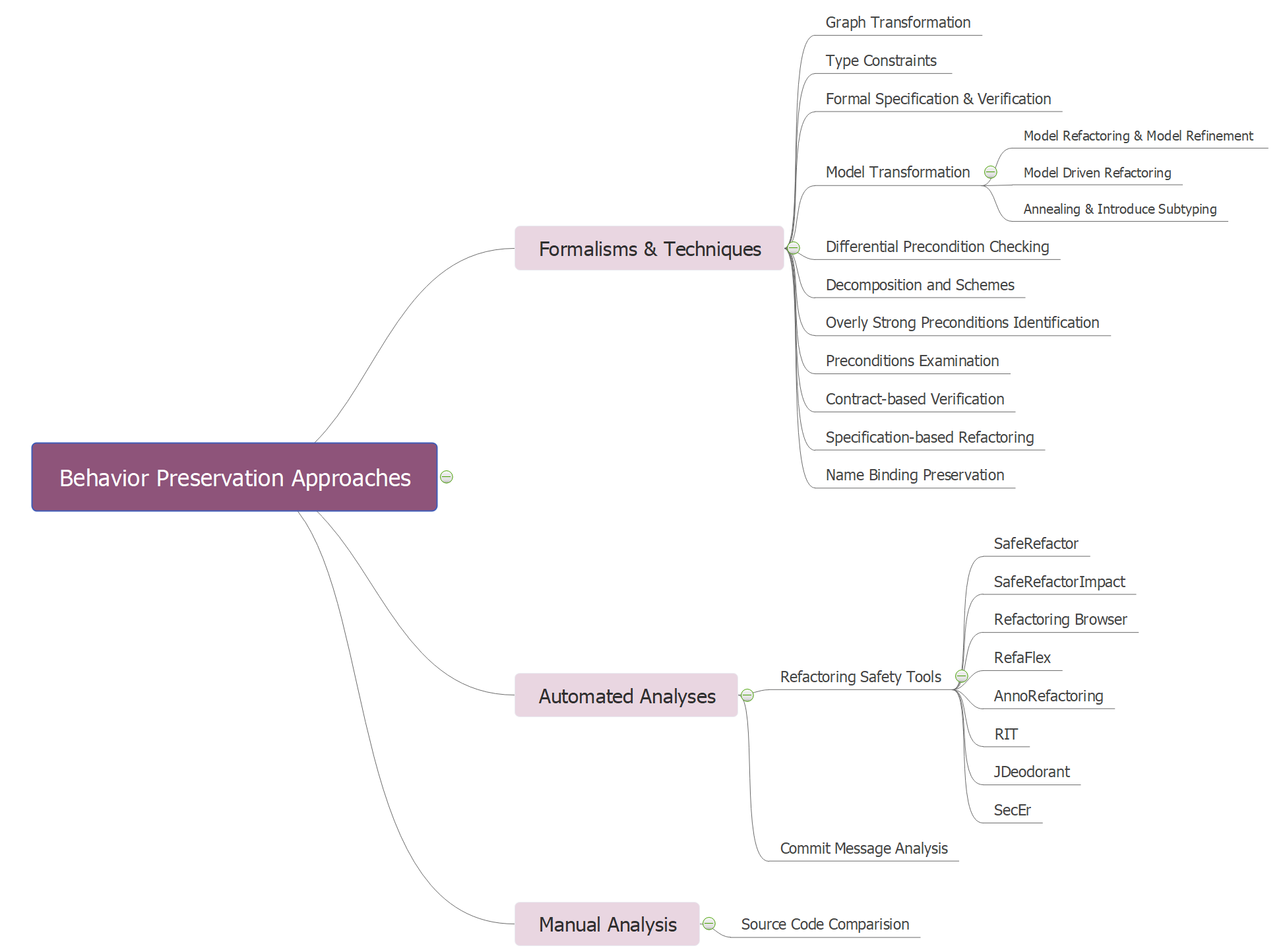}   
    \caption{Behavior Preservation Approaches.}
    \label{fig:Behavior_Preservation_Approaches}
\end{figure*}
\begin{sidewaysfigure}
\centering
\begin{subfigure}{10cm}
\centering\includegraphics[width=10cm]{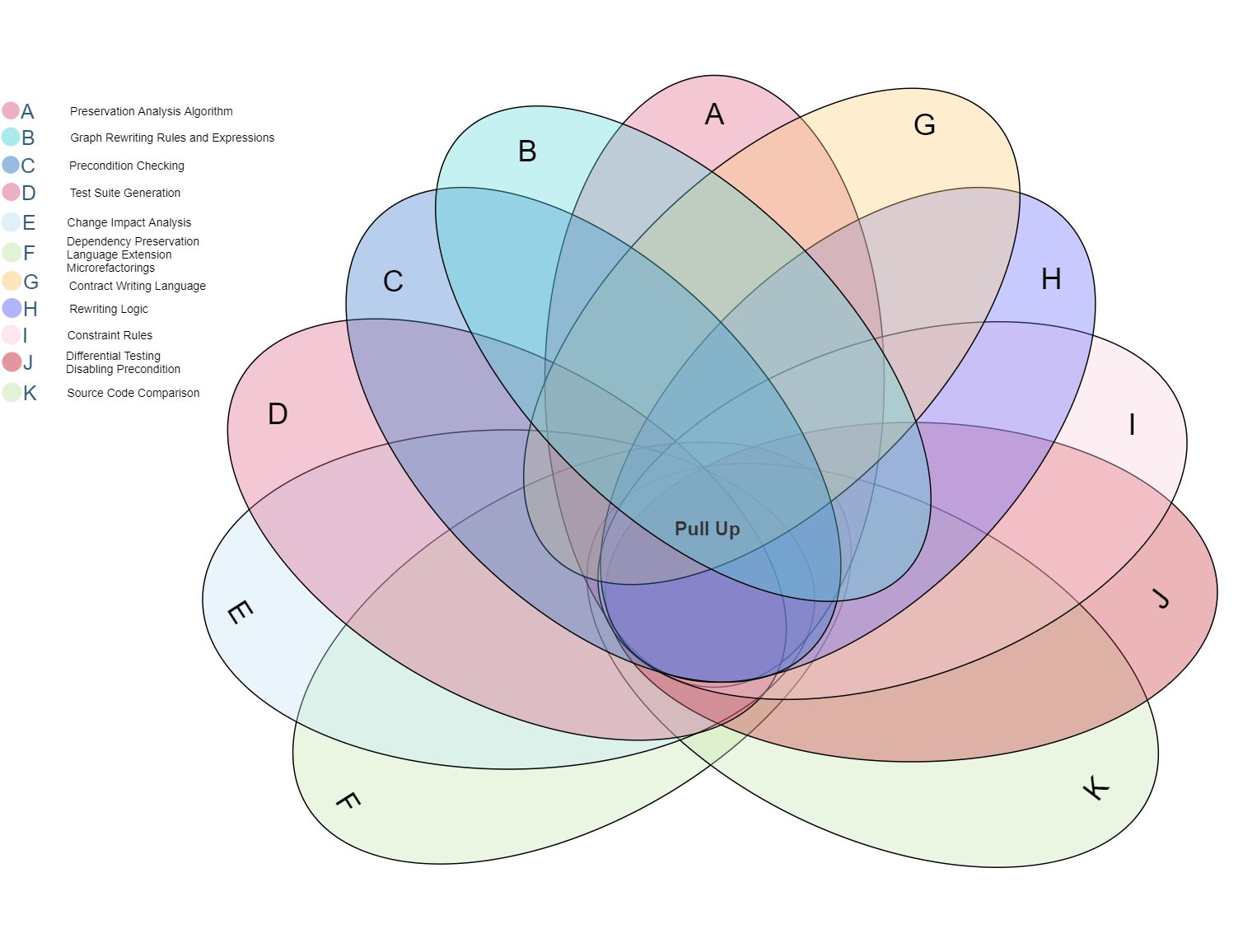}
\caption{Pull up-related Operations}
\label{BP:PullUp}
\end{subfigure}%
\begin{subfigure}{10cm}
\centering\includegraphics[width=10cm]{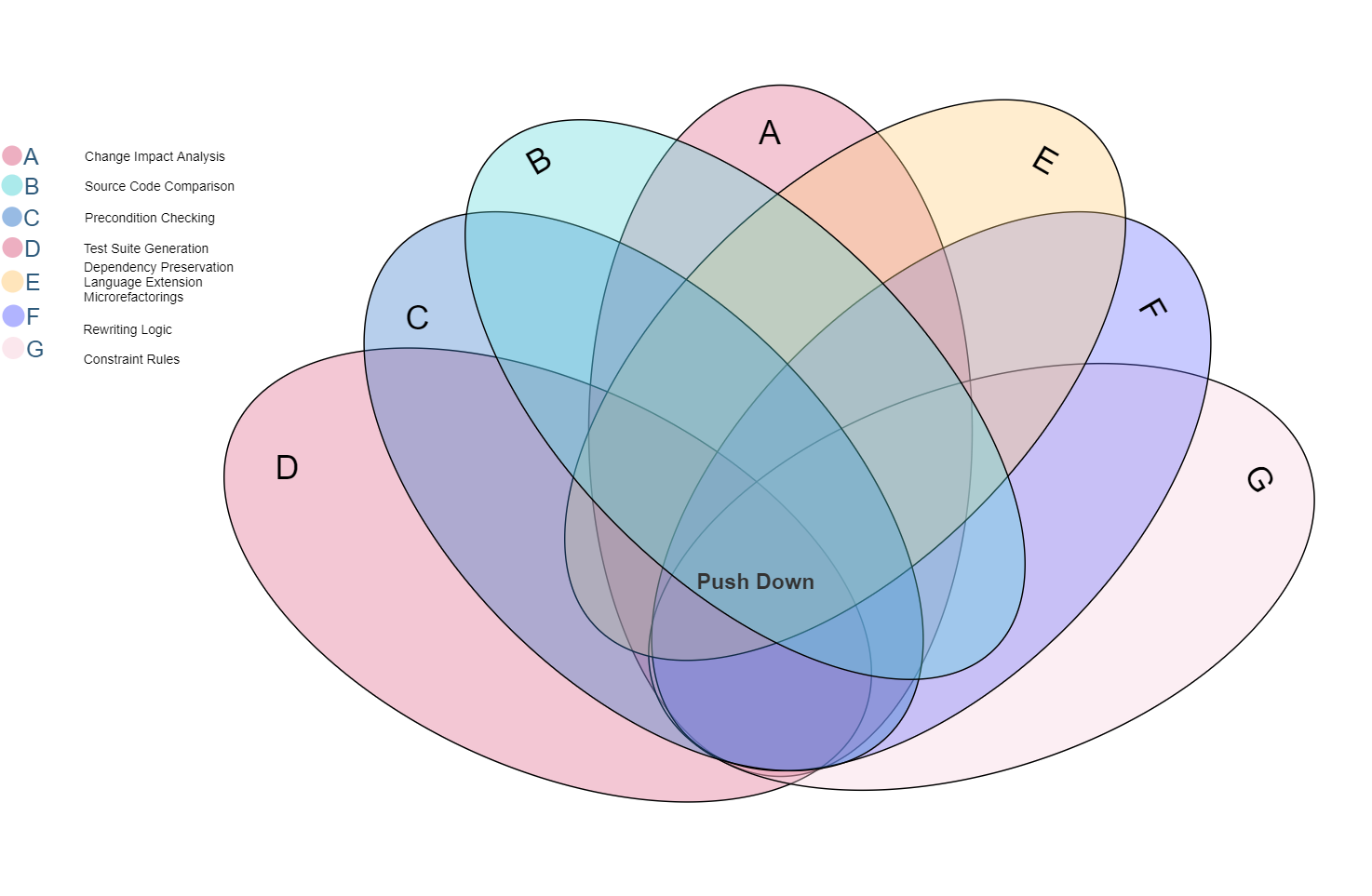}
\caption{Push Down-related Operations}
\label{BP:PushDown}
\end{subfigure}%

\begin{subfigure}{10cm}
\centering\includegraphics[width=10cm]{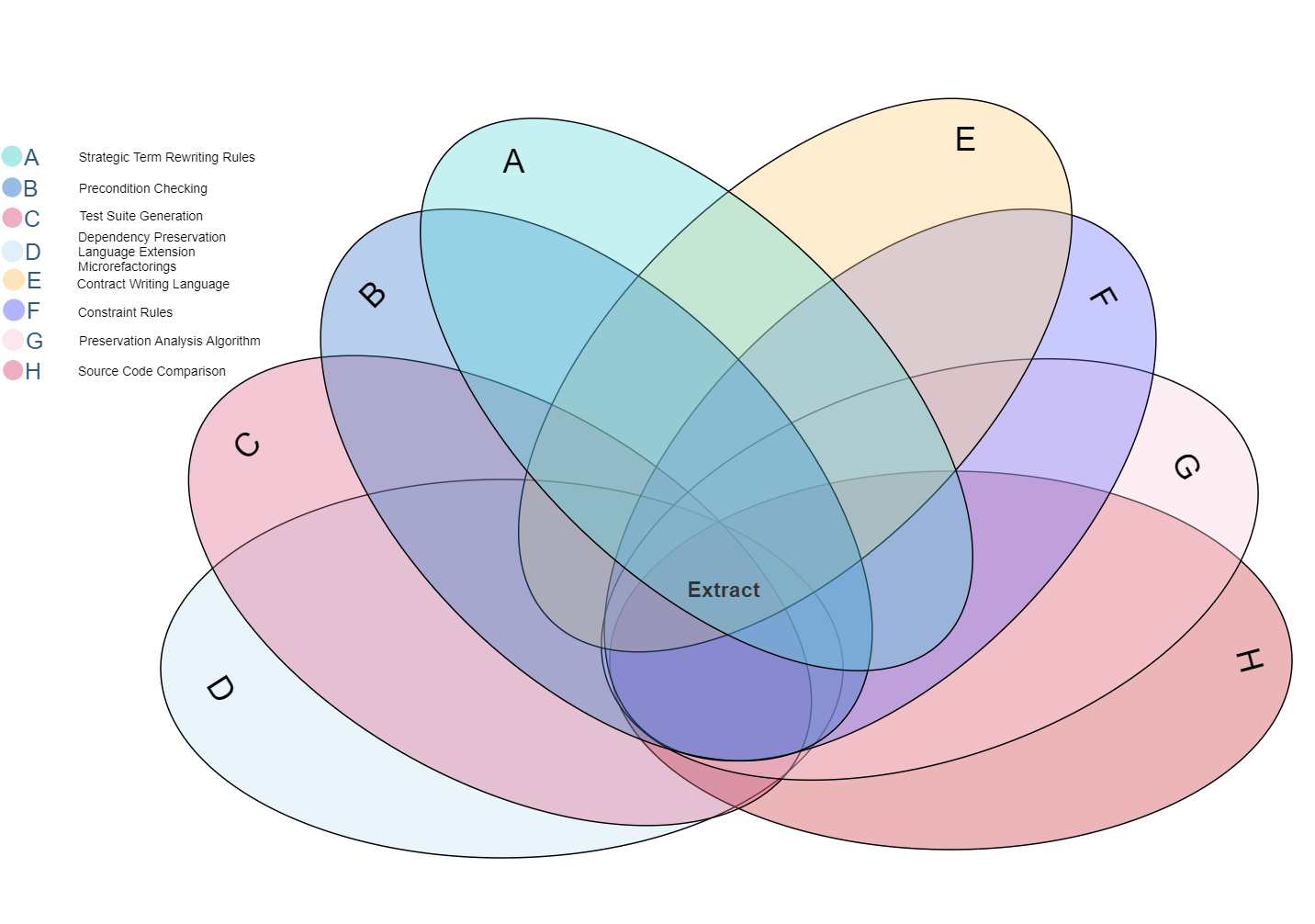}
\caption{Extract-related Operations}
\label{BP:Extract}
\end{subfigure}%
\begin{subfigure}{8cm}
\centering\includegraphics[width=10cm]{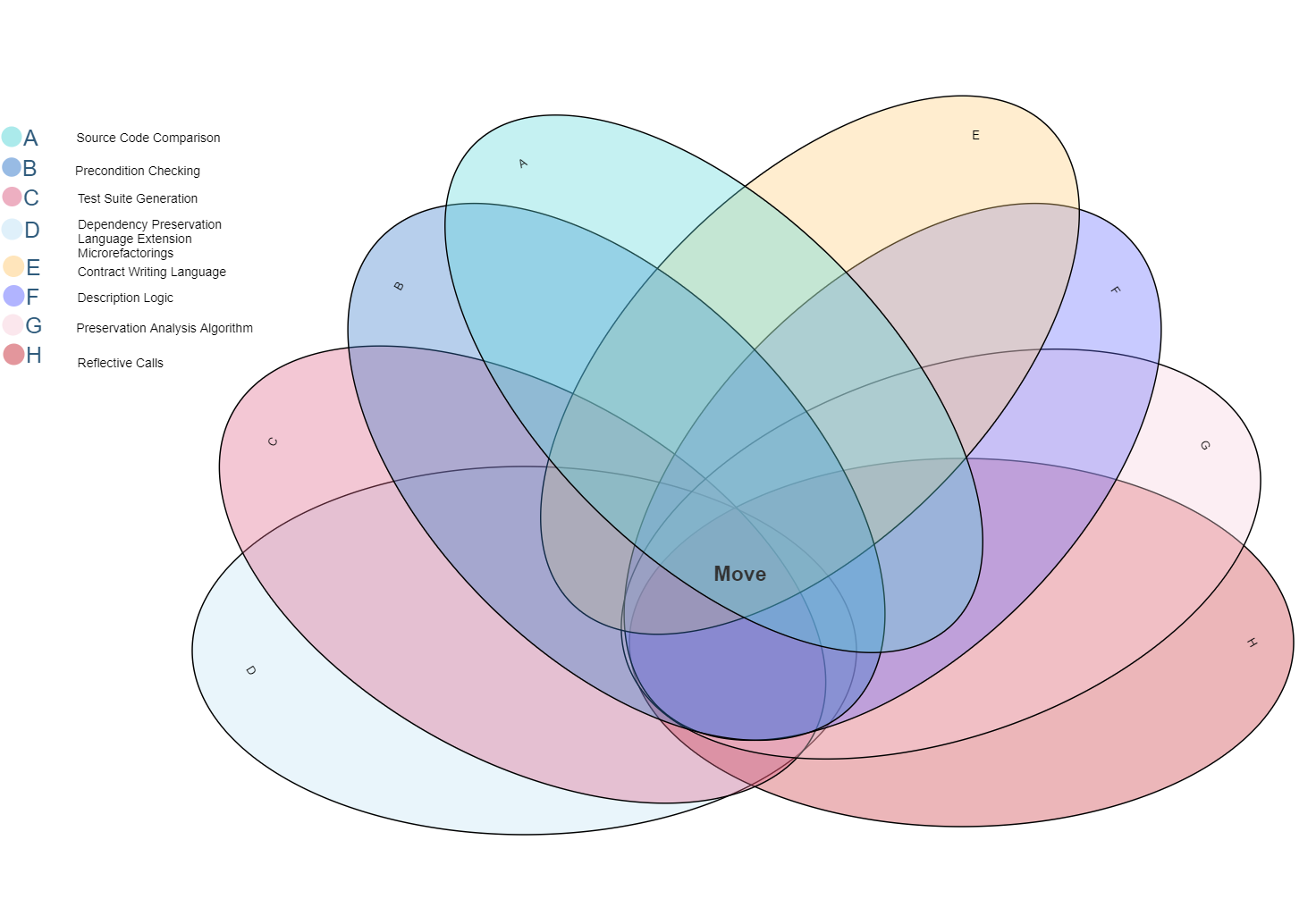}
\caption{Move-related Operations}
\label{BP:Move}
\end{subfigure}%

\caption{Behavior Preservation Strategies and the Evaluated Refactoring Operations.}
\label{overlap_refactoring_1}
\end{sidewaysfigure}
\begin{sidewaysfigure}
\centering
\begin{subfigure}{10cm}
\centering\includegraphics[width=10cm]{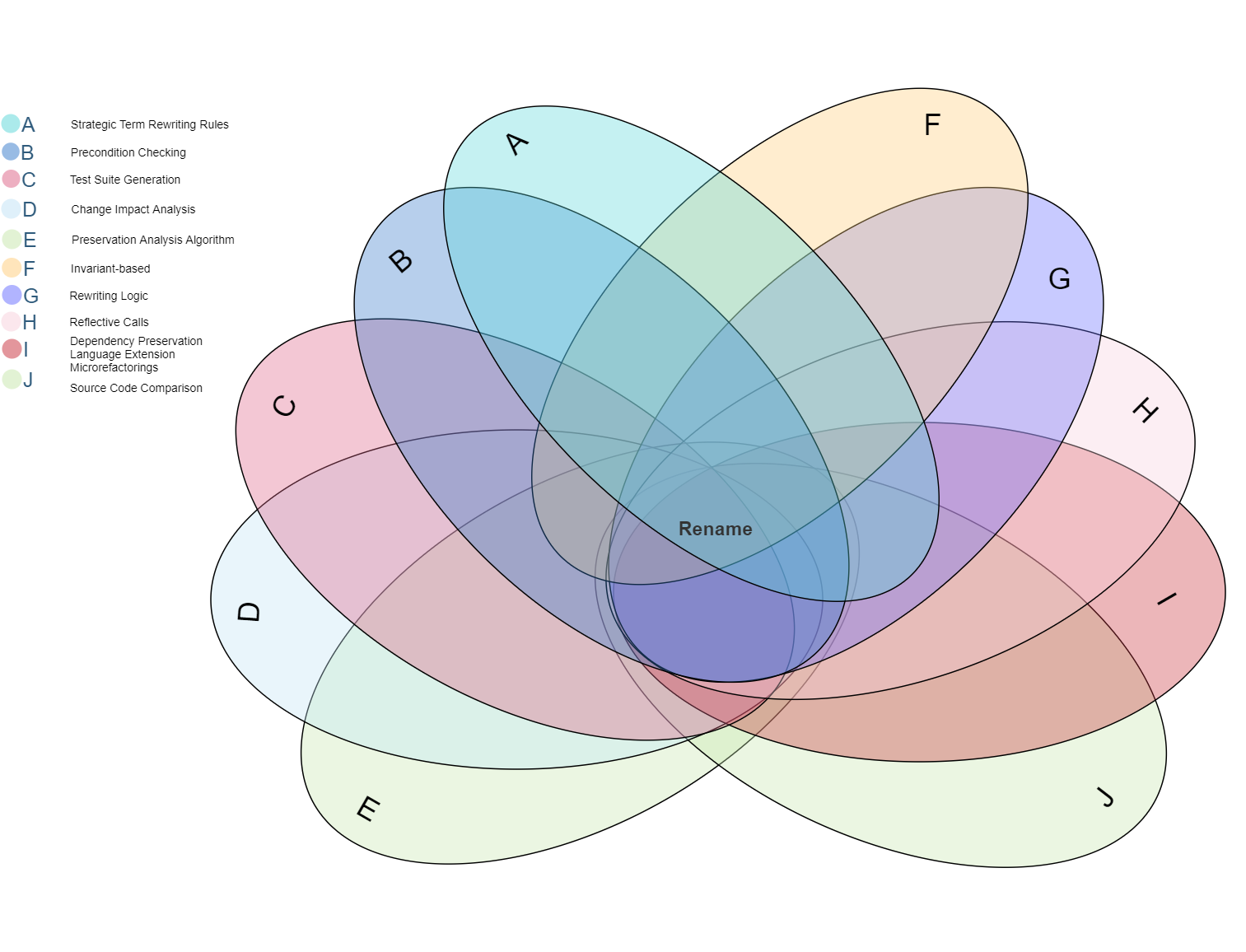}
\caption{Rename-related Operations}
\label{BP:Rename}
\end{subfigure}%
\begin{subfigure}{10cm}
\centering\includegraphics[width=10cm]{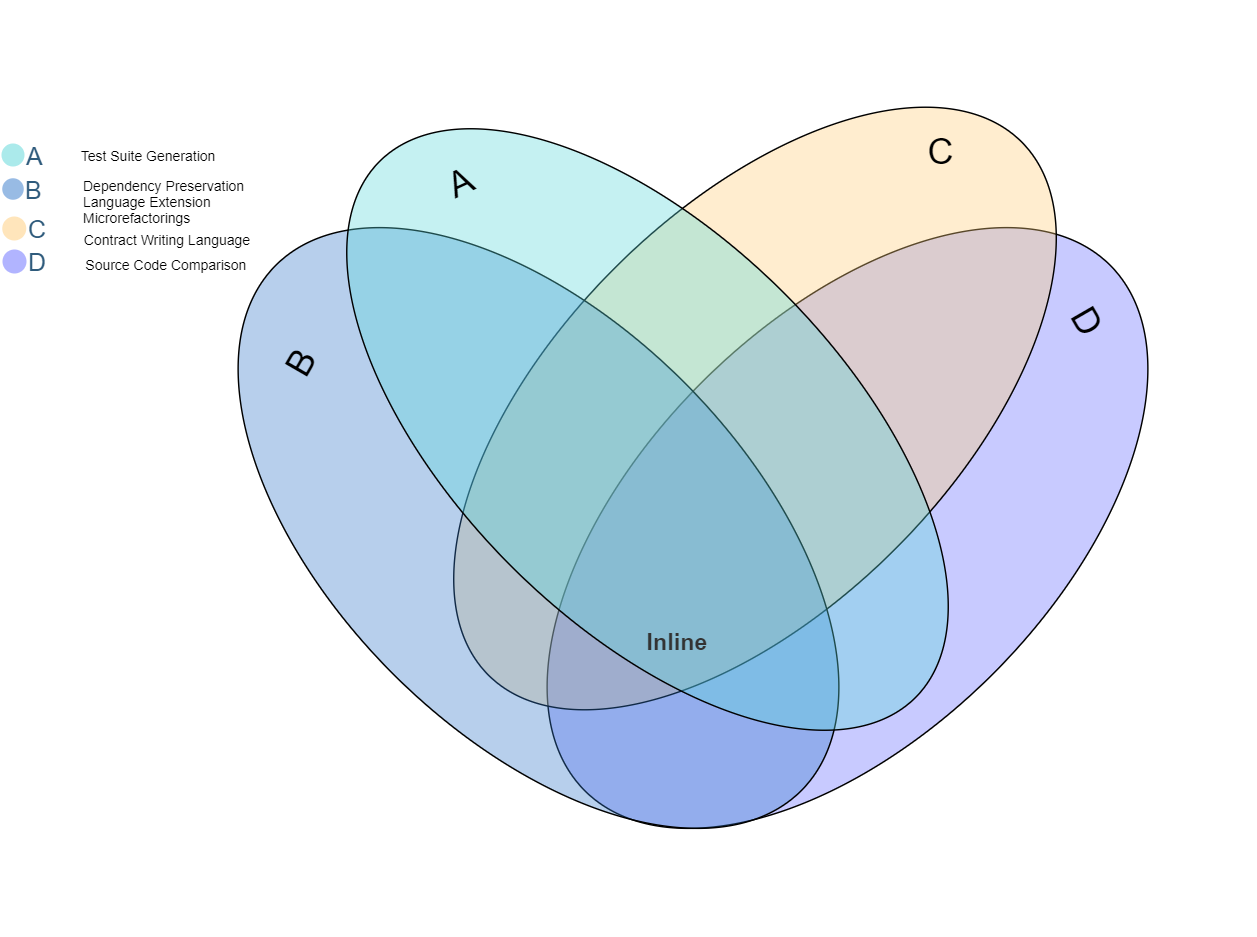}
\caption{Inline-related Operations}
\label{BP:Inline}
\end{subfigure}%


\begin{subfigure}{10cm}
\centering\includegraphics[width=10cm]{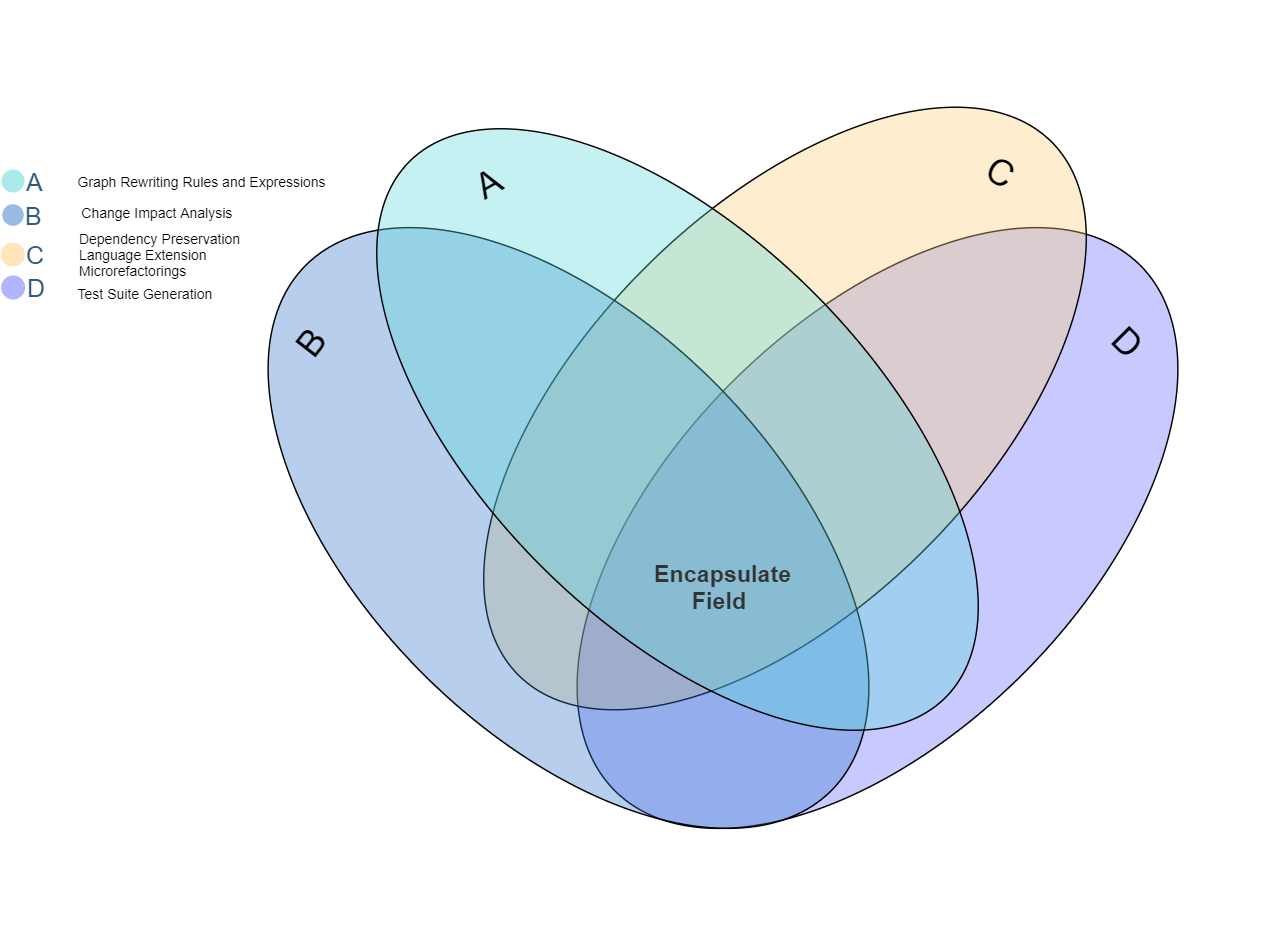}
\caption{Encapsulate Field-related Operations}
\label{BP:Encapsulation}
\end{subfigure}%
\caption{Behavior Preservation Strategies and the Evaluated Refactoring Operations (Cont).}
\label{overlap_refactoring_2}
\end{sidewaysfigure}

All of the accepted literature publications have reported an approach to preserving the behavior. The approaches vary between using formalisms, applying techniques, developing automatic refactoring safety tools, and performing a manual analysis of the source code. We decided to cluster these approaches as follows: (1) refactoring formalisms and techniques, (2) automated analyses, and (3) manual analysis.  Formalism and technique is any behavior preservation approach proposed using a technique or specification. It is not necessarily to be incorporated with a refactoring engine. Automated analysis is any behavior preservation approach that is proposed by incorporating it with a refactoring engine to automate the process. Additionally, we classify the reported approaches into fourteen subcategories. The designed schema for classifying these approaches is depicted in Figure~\ref{fig:Behavior_Preservation_Approaches}. We consider the three already-mentioned classifications as a starting point of the schema, and then classify the reported approaches to each of these classifications. Each PS can belong to one or more subcategories. A detailed overview of these classifications is shown in Table~\ref{Table:Related_Work_Approach}. Figures \ref{overlap_refactoring_1} and \ref{overlap_refactoring_2} depict refactoring operations that are overlapped between multiple strategies. Due to the space constraints, we only show the popular refactoring operations used in the literature, namely, \textit{Pull up}, \textit{Push down}, \textit{Extract}, \textit{Move}, \textit{Rename}, \textit{Inline}, and \textit{Encapsulate Field}. As can be seen, these popular refactorings are evaluated using multiple strategies. More details about the overlapped refactorings can be found in our extension package \footnote{https://smilevo.github.io/self-affirmed-refactoring/}. Detailed descriptions of the approaches are described below.

\begin{table*}[ht]
\begin{center}
\caption{Behavior Preservation Approaches and its Strategies in Related Work.}
\label{Table:Related_Work_Approach}
\begin{adjustbox}{width=1.0\textwidth,center}
\begin{tabular}{lclll}\hline
\toprule
\bfseries Study & \bfseries Year & \bfseries Approach & \bfseries Strategy \\
\midrule
Roberts et al. \cite{roberts1997refactoring} & 1997 & Refactoring Safety Tool & Precondition Checking  \\
Mens et al. \cite{Mens2003FormalisingRW} & 2003 & Graph Transformation & Graph Rewriting Rules \& Expressions  \\
Tip et al. \cite{Tip:2003:RGU:949305.949308} \cite{tip2011refactoring} & 2003,2011 & Type Constraints & Constraint Rules  \\
Garrido and Meseguer \cite{4026866} & 2006 & Formal Specification \& Verification & Rewriting Logic  \\
Straeten et al. \cite{VanDerStraeten2007} & 2007 & Model Transformation & Description Logic  \\
Massoni et al. \cite{Massoni:2008:FMP:1792838.1792873} & 2008 & Model Transformation  & Laws of Programming  \\
Soares et al. \cite{article} \cite{soares2009generating} \cite{5440166}\cite{soares2011making} & 2009,2010,2011 & Refactoring Safety Tool  & Test Suite Generation & \\
Ubayashi et al. \cite{ubayashi2008contract} & 2008 & Contract-based Verification & Contract  Writing Language  \\
Sch{\"a}fer et al. \cite{schafer2008sound} & 2008 & Naming Binding Preservation & Invariant-based  \\
Tsantalis and Chatzigeorgiou \cite{4752842} & 2009 & Precondition Examination & Precondition Checking \\
Sch{\"a}fer and Moor \cite{schafer2010specifying} & 2010 & Specification-based Refactoring & Dependency Preservation \\
& & &  Language Extension\\
& & & Microrefactorings \\
Tsantalis and Chatzigeorgiou \cite{tsantalis2010identification} & 2010 & Refactoring Safety Tool & Precondition Checking  \\
Overbey and Johnson \cite{6100067} & 2011 & Differential Precondition Checking & Preservation Analysis Algorithm      \\ 
Soares et al. \cite{6080784}, Mongiovi et al. \cite{7898404} & 2011,2017 & Overly Strong Preconditions Identification & Differential Testing\\
& & &  Disabling Preconditions  \\

Jonge and Visser \cite{de2012language} & 2012 & Name Binding Preservation &  Invariant-based \\
Noguera et al. \cite{noguera2012refactoring} & 2012 & Refactoring Safety Tool & Annotation-aware  \\
Thies and Bodden \cite{thies2012refaflex} &  2012 & Refactoring Safety Tool & Reflective Calls  \\
Soares et al. \cite{SOARES20131006} & 2013 & Refactoring Safety Tool & Test Suite Generation   \\
& & Commit Message Analysis & Keywords-based Search  \\
& & Manual Analysis & Source Code Comparison \\
Soares et al. \cite{article}, Mongiovi et al. \cite{Mongiovi:2014:MRS:2664678.2664813} & 2009,2014 & Refactoring Safety Tool & Change Impact Analysis  \\
Najaf et al. \cite{najafi2016set} & 2016 & Annealing \& Introduce Subtyping & UML-B Refactoring Rules \\
Horpácsi et al. \cite{Horpcsi2017TrustworthyRV} & 2017 & Decomposition \& Schemes & Strategic Term Rewriting Rules  \\
Chen et al. \cite{chen2018improving} & 2018 & Refactoring Safety Tool & Test Suite  Generation  \\
Insa et al. \cite{insa2018behaviour} & 2018 & Refactoring Safety Tool & Test Suite Generation \\

\bottomrule
\end{tabular}
\end{adjustbox}
\end{center}
\end{table*}

\subsubsection{Refactoring Formalisms and Techniques}
\label{sec:Refactoring Formalisms and Techniques}

This section demonstrates an example that shows some aspects of the Formalisms and Techniques behavior preservation approach. Consider the class \texttt{Employee} and its subclass \texttt{Salesman}. Class \texttt{Employee} declares the \texttt{getName}, \texttt{getSalary}, \texttt{yearlySalary} methods, and 
 Class \texttt{Salesman} declares methods \texttt{setSSN}, \texttt{getSSN}, \texttt{getFullName}, \texttt{getSalary}, \texttt{getSomething}, \texttt{toString}, \texttt{yearlySalary}, \texttt{yearlySalaryIncrease}, \texttt{displayYearlySalaryIncrease}, \texttt{test1}, and \texttt{test2}. Suppose we use generalization-related refactorings (i.e., \textit{Pull Up Attribute} and \textit{Pull Up Method}) to demonstrate this approach. In that case, we notice that one of the strategies listed in Table \ref{Table:Related_Work_Approach} (i.e., preconditions) needs to be checked before performing refactoring, as follows:
\begin{itemize}
\item Methods \texttt{setSSN}, \texttt{getSSN}, and field \texttt{ssn} can be pulled up from class \texttt{Salesman} into class \texttt{Employee} without affecting program behavior.
\item Method \texttt{yearlySalary} cannot be pulled up into class \texttt{Employee} because class \texttt{Employee} has a method with the same signature defined.
\item If method \texttt{toString} is pulled up into superclass, there is no compilation error introduced but the program is behaviorally changed. This is because the call \texttt{s.toString()} dispatches to a different implementation of the method \texttt{toString()}.
\item Method \texttt{displayYearlySalaryIncrease} cannot be pulled up without pulling up \texttt{yearlySalaryIncrease()} because \texttt{yearlySalaryIncrease()} is not declared in class \texttt{Employee}.

\end{itemize}

Some aspects of refactoring formalisms and techniques include displaying the violation of refactoring preconditions. For instance, refactoring tools that display violations should: (1) not take longer than a manual refactoring, (2) indicate all locations of precondition violation, (3) show violated preconditions at once, and (4) display the violation relationally.

\paragraph{Graph Transformation}
Existing refactoring tools lack solid specification of the refactoring procedures. Current specifications are defined by examples or by including assertions (pre/postconditions) that are mostly language-specific. To increase the reliability of these tools, a formal model is required to support software refactoring which can be expressed by a graph transformation. These formal models should: be a language-independent representation of the source code, preserve certain program properties and include formal analysis of the assertion to ensure the completeness of the specification. Further, since refactoring tools represent the source code by abstract syntax tree and any refactoring activity is supposed to change that graph, there is a need to have a formal specification for refactoring that corresponds with a number of graph rewriting rules \cite{Mens2003FormalisingRW}. 

In a study survey, Mens and Tourwe \cite{1265817} summarize these formal properties by showing the correspondence between refactoring and graph transformation as shown in Table~\ref{Table:Formal_Properties_of_Graph_Transformation.}.

\begin{table}[h]
\begin{center}
\caption{Formal Properties of Graph Transformation (Extracted from Mens and Tourwe \cite{1265817}).}
\label{Table:Formal_Properties_of_Graph_Transformation.}
\begin{tabular}{lll}
\toprule
\bfseries  Refactoring  & \bfseries Graph Transformation \\
\midrule
software artifact &   graph\\ 
refactoring & graph production\\ 
composite refactoring & composition of graph productions \\ 
refactoring application & graph transformation\\ 
refactoring precondition & application precondition  \\ 
refactoring postcondition & application postcondition \\ 
\bottomrule
\end{tabular}
\end{center}
\end{table}

\paragraph{Type Constraints}
Tip et al. \cite{Tip:2003:RGU:949305.949308} \cite{tip2011refactoring} propose using type constraints that depend on interprocedural relationships between variable types. A constraint variable can be one of the following: (1) type of constant, (2) type of expression, (3) type of declared method, and (4) type of declared field. To ensure the preservation of program behavior, each refactoring should be associated with a set of preconditions  that must be satisfied. Type constraints mechanism verifies the preconditions and determines the source code that is allowed to be modified.

\paragraph{Formal Specification and Verification}
Due to the lack of accurate specification of the preconditions and lack of proof of the correctness of the refactoring tools, Garrido and Meseguer \cite{4026866} introduce an equational (rewrite logic) semantics-based approach that fulfills two essential goals: (1) formally specifying Java refactorings, and (2) proving behavior-preserving of refactorings with respect to the language's formal semantics. The implementation of this approach is based on the rewrite logic executable semantics of Java refactorings in Maude language. They show the Maude specification of push down method, pull up field, and rename temporary Java refactorings along with providing a mathematical proof of the correctness for two of those refactoring operations. They particularly prove that these refactorings preserve the program behavior with reference to the formal Java semantics. 

Consider the formal specification of \textit{Pull Up Attribute} defined in \cite{4026866}. By applying this refactoring operation on field \texttt{ssn} to move the field to the class \texttt{Employee}, the following preconditions must hold in order for transformation to be carried out successfully. 
\begin{itemize}
\item There is a class named \texttt{Employee}.
\item Class \texttt{Employee} has at least one subclass.
\item Class \texttt{Employee} does not define the field \texttt{ssn}.
\item Subclass of \texttt{Employee} defines the field \texttt{ssn}.
\end{itemize}

These preconditions are checked by \texttt{preconditionsPullUpFieldHold} operation and applied by operation \texttt{applyPullUpField} in the formal specification listed in \cite{4026866}.

\paragraph{Model Transformation}
\subparagraph{Model Refactoring and Model Refinement.}
Straeten et al. \cite{VanDerStraeten2007} differentiate between model refactoring and refinement as follows: model refactoring aims at improving the model structure while preserving its behavior, while model refinement aims at providing more detail to an existing model. The main purpose of this formalism is to investigate the relation between the behavior inheritance consistency and behavior preserving properties of a refined model and a refactored model respectively. These model transformation activities are used to manipulate models, and are supported by a practical formalism that detects the behavior consistency between a refined model and a refactored model. This is achieved by the developed plug-in in a UML CASE tool. These two kinds of model transformation need to be complemented by model inconsistency management to avoid any possibility of inconsistency between models after the transformation activity. Straeten et al. \cite{VanDerStraeten2007} used reasoning capabilities of descriptive logics (DLs) \cite{baader2003description} to detect behavioral inconsistencies during model refinement and behavioral preservation violation during model refactoring. In other words, they check the behavior inheritance consistency between superclass and its subclass and then ensure that this behavior consistency is preserved between refactored classes in an inheritance hierarchy.


\subparagraph{Model-Driven Refactoring.}
Some of the popular model-driven development approaches (e.g. Round-trip Engineering) generate changes to programs from a model, which requires manual updates, making the evolution costly \cite{Massoni:2008:FMP:1792838.1792873}. To avoid any manual update activity on the source code, Massoni et al. \cite{Massoni:2008:FMP:1792838.1792873} propose a formal approach to refactor programs in a model-driven manner in a semi-automatic way which guarantees behavior preservation of the target program. The precondition for applying this approach is to guarantee that the source code is in conformity with the object model. In this approach,  Alloy is used as the model transformation system, where each primitive Alloy model transformation from the catalog is associated with a strategy to refactor the program. This results in a program consistent with the refactored program (i.e. behavior preservation). This formal approach is guided by laws of programming that have been proven to be behavior-preserving.

\subparagraph{Annealing and Introduce Subtyping.}

Najafi et al. \cite{najafi2016set} proposed annealing
and introduce subtyping rules as refactoring rules which can improve the design from an abstract specification written in UML-B. These rules are similar to refactoring rules proposed by Fowler et al. \cite{Fowler:1999:RID:311424}. Annealing adds structure to a specification by splitting a class
into two classes (aiming for more fine-grained classes); whereas introducing subtyping
establishes a relationship between classes with many features in common (with
the aim of increasing reusability). The rules are behaviorally preserved as it ensures that any software
design produced will be correct with respect to the original specification.

\paragraph{Differential Precondition Checking}
Due to the importance of setting preconditions to help guarantee behavior-preserving program transformations for automated refactorings, Overbey and Johnson \cite{6100067} propose a technique called Differential Precondition Checking. This technique has the added advantages of being language independent and reusable in a library. It checks for preservation by first analyzing the source code and creating program representation. Before validating user input, it constructs a program graph as a semantic model (i.e. initial model). It then produces a new program representation for the modified source code for the purpose of checking compilability of the refactored program. For this new program representation, it also constructs a derivative semantic model. The following step is to perform preservation analysis by comparing the two semantic models. If the differential precondition checker determines that the transformation is behavior preserving, the modification will be applied to the source code. Otherwise,  the refactoring will not be considered.

By way of illustration, Overbey and Johnson \cite{6100067} show the differences between the traditional precondition checking and the differential checking for \textit{Pull Up Method} refactoring. For the traditional version, the method needs to be moved from subclass to its superclass, replacing all occurrences of superclass with \texttt{this}. Using preservation rule for the differential version, however, this refactoring is composed of two smaller refactoring operations: (1) \textit{Copy Up Method} to move a method to its superclass and replace all occurrences of the superclass with \texttt{this} and (2) \textit{Delete Overriding Duplicate} to delete the original method from the subclass using the preservation rule in \cite{6100067}. 

\paragraph{Decomposition and Schemes}
Some of the previously built complex refactoring tools were solidly developed, but not totally accurate in guaranteeing the correctness of the transformation these tools implement, which have resulted in introducing bugs to the system. In order to solve this problem, Horpácsi et al. \cite{Horpcsi2017TrustworthyRV} propose to decompose complex refactoring transformation into a series of prime refactorings that can also be expressed as instances of refactoring schemes, and then verified based on formal program semantics. This way, the transformations become simple and more easily verified.

\paragraph{Overly Strong Precondition Identification}
Soares et al. \cite{6080784} propose an approach to identify overly strong conditions in refactoring implementations as these conditions may prevent behavior-preserving transformations. This formal specification helps in guaranteeing program preservation. The process of checking overly strong conditions begins with an automatic generation of Java programs as test inputs using a program generator called JDolly. For each program generated by JDolly, the same refactoring is applied by using three different refactoring implementations (i.e., Eclipse, JRRT, NetBeans). Then, the outputs of the refactoring implementations are compared. To evaluate whether a transformation is behavior-preserving, SafeRefactor is used to identify behavioral changes in transformation. If one implementation rejects the transformation, and the other implementation accepts it with the conformation from SafeRefactor tool that it is behaviorally preserved,  this is an indication that the first implementation that rejects the transformation contains an overly strong condition. This technique is also called Differential Testing (DT) \cite{article3}. 

For an example of such an overly strong condition, suppose we apply \textit{Rename Method} refactoring to rename method \texttt{getFullName} to \texttt{getName}. If we apply this refactoring using Eclipse, we get the following warning message: \textit{Problem in 'Salesman.java'. The reference to \texttt{getName} will be shadowed by a renamed declaration}. After applying the transformation, the \texttt{test2} method outputs \texttt{John Smith} instead of \texttt{John}. This transformation exposes a behavioral change after ignoring a warning message. Similarly, NetBeans applies the transformation.

By applying this refactoring using JRRT, however,  the transformation preserves behavior. 
JRRT adds a \texttt{super} access to method \texttt{getName} inside \texttt{test2} to ensure that the resulting program 
correctly refactors the source program. 

We notice that Eclipse rejects the transformation, and NetBeans and JRRT apply it with the conformance from SafeRefactor tool that it is behaviorally preserved. Thus, by comparing the results of Eclipse, JRRT, and NetBeans, it indicates that Eclipse has an overly strong condition because it rejects useful behavior preserving transformation. 

In the following study that complements this work, Mongiovi et al. \cite{7898404} propose a new technique called Disabling Preconditions (DP) to detect overly strong preconditions. The process starts with using JDolly as test inputs (Step 1). For each generated program, the refactoring engine is used to apply the transformations. In Step 2, authors collected the messages reported by the refactoring engine about the rejection of certain refactoring transformations. The next step is to manually inspect the code fragments and its related precondition for the purpose of disabling the execution of the precondition (i.e., DP technique). Step 5 involves reapplying the same transformation with a disabled precondition. After ensuring that the refactoring implementation applies the transformation and this transformation is behaviorally preserved according to SafeRefactorImpact, DP technique classifies a precondition as overly strong precondition.

\paragraph{Behavior Preservation Preconditions Examination}

Tsantalis and Chatzigeorgiou \cite{4752842} propose a methodology to preserve the behavior of the code by examining a set of preconditions when applying \textit{Move Method} refactoring. These preconditions should be satisfied in order to avoid behavioral changes. Tsantalis and Chatzigeorgiou \cite{4752842} formally define a set of auxiliary functions that describe behavior preservation preconditions as follows:
\begin{itemize}
\item A class should not inherit a method having a matching signature with the moved method. This action will lead the inherited method to override causing behavioral changes of the target class and its derived one. The moved method needs to be renamed to resolve the issue. 
\item When moving a method, the method should not override an inherited method. The original method should be kept as delegate to the moved method.
\item When moving a method, the method should have a valid reference to its target class. The moved method can have a reference via its parameters or fields in the original class.
\item When moving a method, the method should not be synchronized. Moving the synchronized method might cause concurrency issues to the original class's objects. 
\end{itemize}

\paragraph{Contract-based Verification}

Ubayashi et al. \cite{ubayashi2008contract} proposed the notion of Refactoring by Contract (RbC) to verify the applied refactoring based on contracts. The contracts in RbC consist of preconditions (i.e., which conditions can be applied), postconditions (i.e., which conditions should be verified after refactoring), and invariants (i.e., what conditions refactoring should preserve). This contract is described in Contract Writing Language (COW), to describe a predicate based on first-order logic. Another study by Dao et al.  \cite{dao2016approach} verified the execution preservation of refactored program which is performed by design patterns. The authors proposed consistent rules (i.e., pre/postconditions) to verify if the execution of the original program and refactored one is preserved the same constraints in the evolution process. 

\paragraph{Specification-based Refactoring}

Because the precondition-based approach is hard to maintain, Sch{\"a}fer and Moor \cite{schafer2010specifying} presented an approach that is based on the concepts of dependency preservation, language extensions, and microrefactorings. The authors pointed out that the approach is powerful enough to provide high-level and precise specifications of
many of the refactorings. The author validated their implementation on Eclipse’s own extensive test suite.

\paragraph{Name Binding Preservation}

Since name binding associates identifiers with program code elements, it forms  a semantic concern that should be preserved by refactorings.  Schäfer et al. \cite{schafer2008sound} pointed out the two limitations in current refactoring tools: (1) too weak preconditions that lead to unsoundness where names do not bind to the correct declarations
after renaming, and (2) too strong preconditions that prevent renaming of certain programs. The authors proposed an invariant-based approach for name binding. In another study,
Jonge and Visser \cite{de2012language} focused on the behavior preservation of static name bindings by implementing a name binding preservation criterion that reuses the name analysis defined in the compiler front-end. This way, even when the language evolves, the semantics assumed by the refactoring tool is guaranteed to be consistent with the semantics implemented in the compiler.

\subsubsection{Automated Analyses}
\label{sec:Automated Analyses}
\paragraph{Refactoring Safety Tools}
\subparagraph{SafeRefactor.}
With the emerging use of refactoring tools, evidence show that these tools do not always preserve behavior since they may lead to erroneous transformations \cite{5440166} \cite{article}. In order to avoid such refactoring errors, Soares et al. \cite{article} developed a tool named SafeRefactor to check refactoring safety. It generally works by identifying behavioral changes in transformations of sequential Java programs and then generating a test suite for capturing unexpected behavioral changes. This process splits into five major sequential steps. After receiving two versions of the program as an input, a static analysis detects common methods in both the source and target programs (step 1). The next step involves generating unit tests for methods identified in step 1 to pinpoint the incorrectly performed refactorings. In step 3, the tool executes the generated test suite on the source program and then runs the same test suite on the target program (step 4). The last step validates whether the transformation introduces behavioral changes: If a test runs successfully in one program and fails in the other, the tool identifies a behavioral change. Otherwise, no behavioral changes will be detected and the transformation is behaviorally preserved.

\subparagraph{SafeRefactorImpact.}
The SafeRefactor tool has been extended, and includes AspectJ support \cite{soares2011making}, uses change impact analyzer called SAFIRA, and generates a test suite only for the methods impacted by the transformation \cite{mongiovi2011safira,Mongiovi:2014:MRS:2664678.2664813}. SafeRefactor was renamed SafeRefactorImpact in \cite{Mongiovi:2014:MRS:2664678.2664813}. This tool works by: (1) comparing the original and modified programs to identify entities (methods) impacted by the change, (2) performing a  change impact analysis technique for the impacted methods in both program versions identifying methods that can be behaviourally changed after the transformation, (3) generating a test suite for the common methods identified in the previous step, (4) executing the test suite before and after the transformation, and (5) evaluating the results of the transformation to determine whether the transformation is behavior preserving. 

Mongiovi et al. compare these tools in \cite{Mongiovi:2014:MRS:2664678.2664813} with respect to several criteria: program correctness, performance, number of methods considered for test generation, change coverage, and relevant tests generated. Their findings show that the extended tool generates better results.


\subparagraph{Refactoring Browser.}

Roberts et al. \cite{roberts1997refactoring} developed a tool called Refactoring Browser, which uses a set of preconditions to ensure a safe and a correct refactoring implementation. This tool was designed solely to automate refactorings for the Smalltalk language. The tool is used regression testing to assure that refactorings indeed do not alter the program’s behavior. In order to preserve the behavior of the program, each refactoring is associated with a reused set of preconditions that must be checked by the compilation framework in VisualWorks. For instance, to successfully implement \textit{Add Method} refactoring, the method name should not conflict with a method defined in the class.

\subparagraph{RefaFlex.}

Since reflective calls are the threats to the validity of refactorings, Thies and Bodden \cite{thies2012refaflex} proposed RefaFlex Eclipse plugin tool for reflective Java programs to ensure refactoring safety. The tool used dynamic analysis to log reflective calls during test runs and then utilized the information to prevent the execution of refactorings that could alter the program's behavior.

\subparagraph{AnnoRefactoring.}

During the condition checking phase of the performed refactoring, annotations can break the behavior preservation as the annotation's restriction can be ignored which no longer guarantee the preservation of the domain-specific mappings. To address this problem, Noguera et al. \cite{noguera2012refactoring} developed an annotation-aware refactoring tool that is integrated with Eclipse to document the domain dependencies that the annotations introduce. Instead of augmenting the refactoring preconditions with the annotation behavior specification, the authors implemented the annotation behavior preservation as post-conditions. Since the refactoring-aware annotation is considered as dependency preservation problem, checking whether the dependencies were maintained after refactoring is crucial.

\subparagraph{RIT.}

Chen et al. \cite{chen2018improving} proposed an Eclipse plugin tool named Refactoring Investigation and Testing (RIT) in order to validate refactoring changes and ensure that the changes behave as intended. For each set of identified refactoring changes, The tool analyzed the original and edited version of the programs, and then detect tests whose behavior might have been modified by refactoring edits. The developed tool helps developers detect refactoring edits responsible for test failures.

\subparagraph{JDeodorant.}

Tsantalis and Chatzigeorgiou \cite{tsantalis2010identification} proposed a technique, implemented as an Eclipse plugin, that extracts refactoring suggestions introducing polymorphism to ensure the behavior preservation based on the examination of a set of preconditions. This technique helps with eliminating the state-checking problem that impacts code quality, and its maintenance requires significant effort.

\subparagraph{SecEr.}

Insa et al. \cite{insa2018behaviour} developed a Sofware Evolution Control for Erlang (SecEr) tool to automatically obtain a test suite that specifically focused on comparing the old and new versions of the code to check the behaviour preservation. Differently from SAFIRA \cite{mongiovi2011safira,Mongiovi:2014:MRS:2664678.2664813} that focused on refactoring as a cause of the change, SecEr is independent of the cause of the changes, being
able to analyze the effects of any change in the code regardless of its structure. All the analyses
performed by the tool are transparent to the user except that it requires user intervention when identifying
point of interests in both the old and the new versions of the program.

\paragraph{Commit Message Analysis}
One of the approaches to analyze refactoring activity on software repositories is by analyzing commit messages. Ratzinger \cite{citeulike:2881658} and Ratzinger et al. \cite{Ratzinger:2008:RRS:1370750.1370759} propose this simple and fast approach to detect refactoring activity between a pair of program versions to determine whether a transformation is behavior preserving. They identified refactorings based on a set of keywords existing in the commit message. In particular, they focus on the following terms in their search approach: \textit{refactor, restruct, clean, not used, unused, reformat, import, remove, replace, split, reorg, rename, and move}. 

Few commit messages containing some of these terms are extracted from the Hadoop\footnote{https://github.com/apache/hadoop} project, as illustrated in the following comments: 

\begin{displayquote}

\par \say{\textit{1. HADOOP-9805. Refactor RawLocalFileSystem rename for improved testability. Contributed by Jean-Pierre Matsumoto.}}


\par\say{\textit{2. HDFS-7743. Code cleanup of BlockInfo and rename BlockInfo to BlockInfoContiguous. Contributed by Jing Zhao.}}


\end{displayquote}


\subsubsection{Manual Analysis}
\label{sec:Manual Analysis}
Murphy-Hill et al. \cite{6112738} identifies refactoring activities by manually analyzing and comparing the source code before and after the commit. For example, to check whether each file before and after the commit preserved behavior, evaluators first review the code to understand the syntax and semantic changes and then use diff tool to help them analyze the transformation. After that, they classify the code changes as either refactoring (such as \textit{Move class} or \textit{Inline method}) or non-refactoring (such as \textit{Add null check}). In case of disagreements on whether the applied refactoring changed the behavior, evaluators discussed them until agreement was reached.

Soares et al. \cite{SOARES20131006} compared and evaluated three approaches, namely,  manual analysis, commit message, and dynamic analysis (SafeRefactor approach) to analyze refactorings on open source repositories, in terms of behavioral preservation. They found, in their experiment, that manual analysis shows the best results in the comparison and is considered the most reliable approach in detecting behavior-preserving transformations.


\begin{tcolorbox}
\textit{Summary.} Many behavior preservation approaches have been proposed in the literature. The approaches vary between using formalisms and techniques, developing automatic refactoring safety tools, and performing a manual analysis of the source code. Researchers are biased toward using precondition-based and testing-based approaches although there are other techniques (e.g., graph-based) that have some potential and perhaps it is effective for certain problems that have not yet well-explored. Several possible strategies can be combined to better detect any violation of the program semantics. 
Formalism and technique approaches are mainly precondition-based, graph-based, model-based, and decomposition-based techniques; automated approaches either rely on testing, preconditions, or keywords. Manual approach is comparison-based in which the source code has been compared before and after the commit.
\end{tcolorbox}
\subsection{RQ4: What evaluation methods were used in the PSs to assess the proposed behavior preservation approaches?}
\label{Result:RQ4}

\begin{table*}[h]
\begin{center}
\caption{Evaluation Methods Used by the Primary Studies.}
\label{Table:Evaluation Methods.}
\begin{adjustbox}{width=1.0\textwidth,center}
\begin{tabular}{lll}
\toprule
\multicolumn{1}{l}{\bfseries {Methods}} & \bfseries No. of PSs & \bfseries  PSs  \\
\midrule
Comparison-based & 5 & \cite{6100067} \cite{SOARES20131006} \cite{Mongiovi:2014:MRS:2664678.2664813} \cite{7898404} \cite{insa2018behaviour}   \\ 
Empirical-based & 13 & \cite{Tip:2003:RGU:949305.949308} \cite{article} \cite{soares2009generating} \cite{5440166} \cite{6080784} \cite{soares2011making}  \cite{thies2012refaflex} \cite{tip2011refactoring} \cite{schafer2010specifying} \cite{schafer2008sound}  \cite{de2012language} \cite{chen2018improving} \cite{noguera2012refactoring} \\ 
Formal Specification-based & 7 & \cite{Massoni:2008:FMP:1792838.1792873} \cite{VanDerStraeten2007} \cite{Horpcsi2017TrustworthyRV} \cite{4026866}  \cite{ubayashi2008contract}  \cite{najafi2016set} \cite{schafer2008sound}  \\ 
Qualitative-based & 1 & \cite{4752842}   
  \\  
Independent assessment-based & 2 & \cite{4752842} \cite{tsantalis2010identification} \\
 
\bottomrule
\end{tabular}
\end{adjustbox}
\end{center}
\end{table*}

Except for \cite{Mens2003FormalisingRW} and \cite{roberts1997refactoring}, all of the PSs used certain evaluation methods to validate their approach. We identified five different evaluation method categories. The applied methods include comparing the approach against others \cite{Mongiovi:2014:MRS:2664678.2664813}\cite{7898404}\cite{6100067}\cite{SOARES20131006} \cite{insa2018behaviour}, running an experiment in one or more refactoring transformations \cite{article}\cite{soares2009generating} \cite{5440166} \cite{6080784}\cite{Tip:2003:RGU:949305.949308} \cite{soares2011making}  \cite{thies2012refaflex}  \cite{tip2011refactoring} \cite{schafer2010specifying}  \cite{schafer2008sound} \cite{de2012language} \cite{noguera2012refactoring}, presenting a formal specification for correctness of refactorings \cite{Massoni:2008:FMP:1792838.1792873}\cite{Horpcsi2017TrustworthyRV} \cite{VanDerStraeten2007} \cite{4026866}  \cite{ubayashi2008contract}  \cite{najafi2016set} \cite{schafer2008sound}, using qualitative analysis \cite{4752842}   \cite{chen2018improving}, and independent assessment \cite{4752842} \cite{tsantalis2010identification}. The authors of \cite{Mens2003FormalisingRW}   don't evaluate their approach, but they plan to validate their approach in the future by the following steps: (1) converting code into a graph, (2) applying graph transformation approach to the graph, and (3) verifying the preconditions for two refactoring operations. Table~\ref{Table:Evaluation Methods.} shows the distribution of the PSs over the evaluation methods and the descriptions are detailed below. 

\subsubsection{Comparison-based evaluation}
Regarding the first evaluation method, the authors of the PSs compare their approach to other existing methods. Overbey and Johnson \cite{6100067} evaluate their approach in three refactoring tools from two different perspectives: the expressivity of the preservation specifications and the performance of differential precondition checking approach compared to a traditional one. Mongiovi et al. \cite{Mongiovi:2014:MRS:2664678.2664813} compare SafeRefactorImpact with SafeRefactor in terms of the similarity of the detected behavioral changes, total time to evaluate the transformation, number of impacted methods, and the change coverage of the generated test suites. Soares et al. \cite{SOARES20131006} compare the three approaches (i.e., SafeRefactor, commit messages analysis, and manual analysis) in terms of identifying all behavior preservation, correctness of the identified behavior preservation, and accuracy of the obtained results. Mongiovi et al. \cite{7898404}  evaluate the approach by comparing bugs detected by Disabling Preconditions (DP) and Differential Testing (DT) techniques. Insa et al.  \cite{insa2018behaviour} compared SecEr with the already available debugging and testing techniques used
when behaviour preservation is checked in an Erlang project.

\subsubsection{Empirical-based evaluation}
For empirical-based evaluation, Soares et al.\cite{5440166} ran the experiment in 24 refactoring transformations using real Java applications and transformations applied by refactoring tools. Soares et al. \cite{soares2009generating} also experimented 16 refactoring cases which successfully detected more than 93\% of errors presented by traditional refactoring tools. Soares et al. \cite{article} evaluate their approach against 9 transformations and the approach did not produce any errors compared to 5 wrongly applied transformations by best refactoring tools. Soares et al. \cite{6080784} assessed their approach by performing an experiment in 27 refactoring operations of three refactoring tools: Eclipse, JRRT, and NetBeans. Tip et al. \cite{Tip:2003:RGU:949305.949308} \cite{tip2011refactoring} implemented only \textit{Extract Interface} refactoring in Eclipse to test the proposed approach. Soares et al. \cite{soares2011making} evaluated the proposed technique in 8 refactorings applied by Eclipse, 23 design patterns, 2 case studies, and 2 JML compilers. Sch{\"a}fer et al. \cite{schafer2008sound} \cite{schafer2010specifying} evaluated the correctness of their refactoring engine in Eclipse test suite. Chen et al. \cite{chen2018improving} applied RIT in 3 Java open source projects that have regression test suites. Jonge and Visser \cite{de2012language} assessed their approach by implementing refactoring for 3 different languages, namely, Mobl, Stratego, and subset of Java. For Mobl and Stratego, they used the existing compilers, whereas for Java subset, they implemented the compiler from scratch. Noguera et al. \cite{noguera2012refactoring} used a prototype extension of the Eclipse IDE's to demonstrate their approach using three annotation libraries: JPA, Aspect5J, and Simple XML. RefaFlex was evaluated in \cite{thies2012refaflex} with 21,524 refactoring runs on 3 open source programs. Their approach prevented 1,358 non behavior preservation transformations.

\subsubsection{Formal specification-based evaluation}
In four PSs, including \cite{Massoni:2008:FMP:1792838.1792873} \cite{Horpcsi2017TrustworthyRV} \cite{VanDerStraeten2007} \cite{4026866}, the approaches were evaluated by formally specifying and verifying the refactoring to ensure that these refactorings are behaviorally preserved. Ubayashi et al. \cite{ubayashi2008contract} evaluated their approach by writing contracts using first-order predicates. Their approach provided good results and most of these contracts can be generated automatically. Najafi et al. \cite{najafi2016set} evaluated their refactoring rules by applying them to an adapted study of
the \textit{Mass Transit Railway System.}

\subsubsection{Qualitative-based evaluation}
In \cite{4752842}, Tsantalis and Chatzigeorgiou assessed their approach using open-source Java projects in four different ways: (1) performing a qualitative analysis of the refactoring suggestions, (2) using software metrics related to coupling and cohesion, (3) having an independent assessment on the refactoring suggestion, and (4) evaluating the efficiency by measuring the computation time with different size of open-source projects. 

\subsubsection{Independent assessment-based evaluation}
In \cite{4752842} and  \cite{tsantalis2010identification}, the proposed approach was evaluated by an independent designer for the system that he developed. The designer provided feedback on the refactoring result from the proposed approach.

\begin{tcolorbox}
\textit{Summary.} With regards to the evaluation methods used in the literature to validate the proposed behavior preservation approaches, PSs used comparison-based, empirical, formal specification-based, quality-based, and independent assessment-based evaluation methods. The majority of PSs empirically evaluate their approaches, and only one study opted for quality-based and independent assessment-based approaches.
\end{tcolorbox}

\section{Discussion and Open Issues}
\label{sec:AnalysisDiscussion}
To ensure that the transformation is behaviorally preserved, we recommend incorporating refactoring tools with the following dimensions:

\begin{itemize}
    \item  Preconditions \& Postconditions \& Invariant:
These properties are used to flag potential violations, such as incompatible signatures in member function redefinition, type-unsafe assignments, or indistinct class and naming \cite{Opdyke:1992:ROF:169783}.  Refactoring tool support needs to determine the number of the preconditions, postconditions, and invariants for each refactoring operation applied by including efficient algorithms for checking these assertions. Although Opdyke proposed a set of refactoring preconditions, there was no formal proof of the correctness of these conditions. Developers should invest into developing more comprehensive refactoring tools by  (1)  adding library containing these assertions to check refactoring so that any refactoring engines for different languages can use this library to test refactoring implementation; and (2) adding formal proofs of the correctness of these assertions to raise the confidence that these set of refactoring help in ensuring that that the transformations preserve the behavior. Additionally, the calculation of pre and postcondition scenarios is time-consuming and error-prone if it is done manually. Future researchers are encouraged to adopt tools to automatically calculate these assertions
and verify the program evolution process.

    \item  Quality Improvement:
In software engineering, maintaining quality is always a top priority.  As development progresses and flaws inevitably begin to emerge, they generate what is known as “code smells”, various indicators that code needs to be refactored or replaced, and can be helpful in identifying problem areas that need to be refactored. Due to the number of design choices, it is challenging to choose the optimal refactorings, maximising the quality of the resulting program while minimizing the cost of behavior preservation transformation. Besides ensuring behavior preservation of the program, it is also advised to check if the resulting program improves the quality of the original program. For instance, the resulting program showcases reusability  and provides trustworthiness by reducing the complexity of the program.

   \item Developer Perception:
Research in preserving the behavior in software refactoring thus far focuses on proposing approaches assuming that the developer's main intention is to perform pure refactoring. Several studies \cite{kim2014empirical,Silva:2016:WWR:2950290.2950305,alomar2019impact,alomar2021we} have been conducted to better understand the motivation behind refactoring (e.g., improving the internal and external structure of the code , removing code smells, etc).  Current approaches have not integrated developers' perception while preserving the behavior of refactoring activities. Researchers should explore developers' insight and experience (e.g., when and how) because they are essential in the behavior preservation process.

    \item    Automated Testing: Some studies [PS7, PS10, PS12, PS13, PS16, PS19, PS23] discussed using testing to ensure behavior preservation but with limited coverage. 
To increase refactoring safety, it is needed to incorporate a solid test suite to the traditional refactoring steps in order to pinpoint non behavior-preservation transformations. That involves generating testing for refactoring applied at different levels of granularity, and taking into account the hierarchy or other object-oriented property.

    \item Tools Availability and Extensibility:  As we noticed in relation to studies [PS1, PS7, PS14, PS21, PS22, PS23, PS28], there is a lack of available tools to support the behavior preservation. Researchers will not be able to adopt  behavior preservation approaches because these tools are not available. As a result, it will make it hard to extend the proposed approaches (e.g., support more refactoring operations, add additional set of preconditions, etc). Additionally, Eclipse plugin tools require user interaction to select projects as inputs to trigger refactorings, which is impractical for a study requiring a high degree of automation since multiple releases of the same project must be imported to Eclipse to check whether the behavior is preserved or not. Further, while some of the current tools warn developers of non-behavior preservation transformations, these tools could be complemented with a compensation transformation that possibly preserve the behavior.  To move the research forward in this area, researchers are advised to implement a full-featured refactoring engine such as integrating  the tools with control version systems like Git or Subversion to easily compare code among several versions and to open source these tools and allow people to replicate and extend them.

    \item   Broader Applicability: Today, a wide variety of refactoring tools automates several aspects of refactoring. However, ensuring the behavior preserving property when building tool-assisted refactoring is challenging. It is acknowledged that refactoring tools should support the following five characteristics: automation, reliability, configurability, coverage, and scalability. Integrating behavior-preserving nature reduces the need to perform testing and debugging. As shown in Figure \ref{fig:Behavior_Preservation_Approaches}, several studies presented many approaches to preserve the behavior. However, we still must understand which approaches are the most effective. While the primary studies proposed refactoring preservation approaches, these approaches should not be language-specific, domain-specific, and refactoring operation-specific. One important research direction is to generalize the behavior preservation approach across multiple languages and multiple domains, and enable semi-automatic formal verification. Researchers are encouraged to explore such interests together with the practice of preserving the behavior in software refactoring.

\end{itemize}

The above mentioned open issues are listed in Table \ref{Table:OpenIssue}. A summary of the findings is reported in Table~\ref{Table:Summary}. We observe that researchers are biased toward certain approaches. As can be seen from the table, researchers extensively used a precondition-based approach. Testing-based is also popular due to the fact that researchers are probably implementing preconditions to test whether the transformation is behaviorally preserved between multiple versions. However, there are other techniques (e.g., graph-based) that have some potential and perhaps it is stronger or effective for certain problems that have not yet explored. Incorporating these specifications in IDE refactoring engines, developers and researchers can revisit existing refactoring tools and extend them.

Recent refactoring research has been taking developer-centric strategies to understand how developers refactor and document their refactorings in practice \cite{zhangpreliminary18,alomar2019can}. Such research has been driven by the rise of several refactoring mining tools \cite{Silva:2016:WWR:2950290.2950305,7962377,tsantalis2018accurate}. Mining the history of previous changes unlocked another dimension of how we should perceive refactoring: Instead of \textit{dictating} how refactoring should be performed and preserved, we can reverse engineer how developers refactor their code and verify the correctness of their operations. Such findings require accurate detection of refactorings, which can be assured by recent studies, as they are reaching a significant precision \cite{tsantalis2018accurate}. Furthermore, the list of mined refactorings has revealed the existence of refactoring types that were absent from studies handling the behavior preservation \cite{tsantalis2020refactoringminer}. 

\begin{table*}[htbp]
\begin{center}
\caption{Open Issues on Behavior Preservation Studies.}
\label{Table:OpenIssue}
\begin{sideways}
\begin{adjustbox}{width=\textheight,totalheight=\textwidth,keepaspectratio}
\begin{tabular}{llll}\hline
\toprule
\bfseries Issue & \bfseries{PSs} & \bfseries Open Issue  \\
\midrule
\textbf{I1} - Assertion  &  &   \multirow{2}{*} {- Researchers can add libraries containing these assertions to test refactoring implementation} \\ 
Precondition  & PS1, PS2, PS3, PS4, PS8, PS11, PS15,  PS17, PS18, PS21, PS26 &   \multirow{2}{*} {- Researchers can add formal proofs of the correctness of these assertions to raise developers' confidence} \\ 
Postcondition & PS2, PS8, PS17, PS21 \\ 
Invariant & PS8, PS9, PS21 \\ \hline
\multirow{2}{*} {\textbf{I2} - Quality Improvement} & \multirow{2}{*} {PS11} & - The studies do not establish an explicit connection between behavior preservation \\ 
&  &  approach and quality, showing there is an opportunity for further studies \\ \hline
\multirow{4}{*} {\textbf{I3} - Developer Perception} & \multirow{4}{*} {N/A} &  - The use of developers' perception and knowledge about refactoring can help to improve \\ 
&  & refactoring process, tools, among other \\ 
&  & - It is essential to evaluate the participation of developers in preserving the behavior, using \\ 
&  & developers' insights and experiences to improve the process \\ \hline
\textbf{I4} - Automated Testing & PS7, PS10, PS12, PS13, PS16, PS19, PS23  & - It is an open theme for researchers to incorporate a solid test suites to test behavior preservation \\ \hline
\multirow{2}{*} {\textbf{I5} - Tool Availability \& Extensibility} &\multirow{2}{*} {PS1, PS7, PS14, PS21, PS22, PS23, PS28} & - There are many opportunities to propose/improve behavior preservation automated tools\\ 
&  & - We need to integrate the tools with control version systems such as Git or Subversion \\ \hline
\multirow{11}{*} {\textbf{I6} - Broader Applicability} & \multirow{11}{*} {N/A} & - We need to explore which approaches are most effective in behavior preservation
\\
&  & - Production of refactoring-agnostic approach \\
& &  - Implementation of language-independence of refactoring schemes \\
&  & - There are many opportunities to research a low explored refactoring operations, most used\\
&  &  refactorings and their relationship with behavior preservation \\
& & - Several possible strategies can be combined to better detect any violation of the program semantics \\
& & - Identification of the appropriate and most reliable evaluation methods to validate the \\
& & future behavior preservation approaches. \\
& & - Refactoring tools could be complemented with a compensation transformation that possibly preserve the behavior. \\
& & - Researchers are encouraged to explore the above-mentioned aspects together with the practice of preserving \\
& & the behavior in software refactoring.\\

\bottomrule
\end{tabular}
\end{adjustbox}
\end{sideways}
\end{center}
\end{table*}
\begin{table*}[!htbp]
\begin{center}
\caption{Summary of Behavior Preservation Approaches in the Primary Studies.}
\label{Table:Summary}
\begin{sideways}
\begin{adjustbox}{width=\textheight,totalheight=\textwidth,keepaspectratio}
\begin{tabular}{llcclccccll}
\toprule
\multirow{2}{*}{\textbf{Study ID}} & \multirow{2}{*}{\textbf{Study}}      & \multicolumn{2}{c}{\textbf{Software Artifact}} & \multirow{2}{*}{\textbf{Language}} & \multirow{2}{*}{\textbf{No. of Ref.}} & \multicolumn{3}{c}{\textbf{Refactoring Classification}}                          & \multirow{2}{*}{\textbf{Approach}} & \multirow{2}{*}{\textbf{Evaluation Method}} \\ 
                                                                      &                             & Code              & Model &  &           & Fowler's catalog & Model ref. & Language-specific &                                                                    \\ 
 \midrule
PS1 & Roberts et al. & Yes & No & Smalltalk & 18 & Yes & No & Yes & Refactoring Safety Tool & Not Mentioned \\
PS2                                                                   & Mens et al.                 & Yes               & Yes &  Not mentioned   & 2         & Yes              & No                & No                & Graph Transformation & Future Validation Tool                                          \\
PS3                                                                   & Tip et al.                  & Yes               & No  & Java & 7             & Yes              & No                & Yes               & Type Constraint & Empirical-based                                                 \\
PS4                                                                   & Garrido \& Meseguer         & Yes               & No & Java & 3                & Yes              & No                & Yes               & Formal Specification \& Verification  & Formal specification-based                    \\
PS5                                                                   & Straeten et al.             & No                & Yes  & UML & 3             & No               & Yes               & No                & Model Transformation &  Formal specification-based                                       \\
PS6                                                                   & Massoni et al.              & No                & Yes  & Alloy              & 7 & No               & Yes               & No                & Model Transformation &  Formal specification-based                                     \\
PS7                                                                   & Soares et al.               & Yes               & No  & Java & 1               & Yes              & No                & No                & Refactoring Safety Tool & Empirical-based                                      \\
PS8 & Ubayashi et al. & Yes & No & AspectJ & 27 & No & No & Yes & Contract-based Verification & Formal specification-based     \\
PS9 & Sch{\"a}fer et al. & Yes & No & Java & 3 & Yes & No & No & Naming Binding Preservation & Formal specification-based   \\
PS10                                                                   & Soares et al.               & Yes               & No    & Java  & 8           & Yes              & No                & Yes               & Refactoring Safety Tool & Empirical-based                                    \\
PS11                                                                   & Tsantalis \& Chatzigeorgiou & Yes               & No    & Java  & 1           & Yes              & No                & No                & Precondition Examination & Quality-based                                  \\
& & & & & & & & &  & Independent assessment-based  \\
PS12 & Sch{\"a}fer \& Moor & Yes & No & Java & 3 & Yes & No & No & Specification-based & Formal specification-based    \\
PS13                                                                   & Soares et al.               & Yes               & No  & Java & 12               & Yes              & No                & Yes               & Refactoring Safety Tool & Empirical-based                                      \\
PS14 & Tsantalis \& Chatzigeorgiou & Yes & No & Java & 1 & Yes & No & No & Refactoring Safety Tool & Quality-based \\
& & & & & & & & &  & Independent assessment-based  \\
PS15 & Tip et al. & Yes & No &  Java & 7             & Yes              & No                & Yes               & Type Constraint & Empirical-based                                                 \\
PS16 & Soares et al. & Yes & No & AspectJ & 7 & Yes & No & Yes & Refactoring Safety Tool & Empirical-based \\
& & & & & & & & &  & Comparison-based  \\
PS17                                                                & Overbey \& Johnson          & Yes               & No & Fortran, PHP, BC  & 18              & Yes              & No                & Yes               & Differential Precondition Checking & Comparison-based                               \\
PS18                                                                  & Soares et al.               & Yes               & No  & Java & 3              & Yes              & No                & No                & Overly Strong Preconditions Identification &  Empirical-based                     \\
PS19                                                                  & Soares et al.               & Yes               & No  &  Java & 36           & Yes              & No                & Yes               & Refactoring Safety Tool & Comparison-based                                        \\
  & & & & & & & & & Commit Message Analysis &  \\
  & & & & & & & & & Manual Analysis & \\
  PS20 & Jonge \& Visser & Yes & No & Java, Stratego, Mobl & 1 & Yes & No & No & Naming Binding Preservation & Quality-based \\
  PS21 & Noguera et al. & Yes & No & Java, AspectJ, XML & Not Mentioned & Yes & No & Yes & Refactoring Safety Tool & Formal specification-based \\
  PS22 & Thies et al. & Yes & No & Java & 6 & Yes & No & No & Refactoring Safety Tool & Empirical-based \\
PS23                                                                  & Mongiovi et al.             & Yes               & No &  Java,AspectJ     & 16            & Yes              & No                & Yes               & Refactoring Safety Tool & Comparison-based                                         \\
PS24 & Najafi et al. & No & Yes & UML & & No & Yes & No & Annealing \& Introduce Subtyping & Formal specification-based  \\
PS25                                                                  & HorpÃacsi                   & Yes               & No   & Erlang   & 6         & No               & No                & Yes               & Decomposition \& Schemes & Formal specification-based                                      \\
PS26                                                                  & Mongiovi et al.             & Yes               & No  & Java & 10             & Yes              & No                & Yes               & Overly Strong Preconditions Identification & Comparison-based                      \\ 
PS27 & Chen et al. & Yes & No & Java  & 5 & Yes & No & No & Refactoring Safety Tool & Empirical-based \\
PS28 & Insa et al. & No & Yes & Erlang & Not Mentioned & No & No & Yes & Refactoring Safety Tool & Empirical-based  \\ 
 \bottomrule
\end{tabular}
\end{adjustbox}
\end{sideways}
\end{center}
\end{table*}

\section{Implication}
\label{sec:Implication}

The main implications of this study are as follows:

\subsection{Implication for practitioners:}

\begin{itemize}
    \item \textbf{Promoting the adoption of behavior preservation approaches in practice.} Due to the growing complexity of software systems, there has been a dramatic increase and industry demand for tools and techniques on software refactoring. Refactoring studies are used in industrial settings and considered objectives beyond improving design to include other non-functional requirements. Thus, challenges to be addressed by refactoring work nowadays  include testing the correctness of applied refactorings. Recent studies (e.g., \cite{vakilian2014alternate,6112738,bogart2020increasing}) show developers under-using automated refactoring tools due to the lack of trust, unawareness, and usability problems. To mitigate this issue, our study reveals several behavior preservation approaches that can be explored to reduce verification effort. For example, developers can use the tool Refactoring Investigation and Testing (RIT) to (1) help them detect refactoring changes responsible for test failures and validating the correctness of the refactored version of the program without the need to rerun the entire regression test suites, and (2) help developers focusing on the long-term management of accidental complexities created by quick design and implementation (e.g., refactoring to reduce technical debt).
    \item  \textbf{Identifying the needed information to the refactored code.} The awareness of such behavior preservation approaches assist programmers in distinguishing precondition violations from warning and advisories without wondering if there are any issues with the applied refactoring. Additionally, it gives programmers an indication of the amount of work required to fix the problem, and so the programmers can determine whether the violation means that the code can be refactored with a few minor changes or not.
\end{itemize}

\subsection{Implication for researchers:}
 \begin{itemize}
     \item \textbf{Developing refactoring tools tuned towards safer refactoring.} As discussed in Section \ref{sec:AnalysisDiscussion}, our study sheds light on a number of desirable properties for refactoring tools (e.g., quality improvement, developer perception, automated testing, etc). Future researchers are encouraged to revisit the existing refactoring tools or build tools that help practitioners have more confidence in using the tools.
     
     \item \textbf{Exploring the potential of combining multiple behavior preservation strategies}. Our study shows that there are some behavior preservation strategies that have been evaluated using single or multiple refactoring operations, and some of these refactorings are applied using multiple strategies. Future researchers are advised to explore the potential of combining several behavior preservation approaches and use the approaches that would be useful in a given context according to a defined set of criteria.
 \end{itemize}
\section{Threats to Validity}
\label{sec:Threats}
In this section, the threats are discussed in the context of four types of threats of validity: internal validity, external validity, construct validity, and conclusion validity.

\noindent{Internal validity:}
Obtaining a representative set of literature publications for this SLM can be viewed as a validity threat due to the search process. To minimize this threat, we followed the SLM guidelines proposed by  \cite{Kitchenham07guidelinesfor,wohlin2014guidelines,petersen2008systematic}. We considered the related search terms and the main terms from research questions to construct the search string and select relevant articles. Further, we followed a four-stage study selection process and applied the inclusion and exclusion criteria in each stage as described in Section~\ref{sec:Research Method}.  Another threat is related to the limitation of the search terms and search engines which might lead to an incomplete set of literature publications. To limit this threat, we used carefully defined keywords and comprehensive academic search engines (i.e., Google Scholar and Scopus) that covers the main publisher venues.

\noindent{External validity:} 
The collected papers contain a significant proportion of academic works which forms an adequate basis for concluding findings that could be useful for academia. However, we cannot claim that the same behavior preservation approaches are used in industry. Also, our findings are mainly within the field of software refactoring. We cannot generalize our results beyond this subject.

\noindent{Construct validity:}
Threats related to the construct validity are the suitability of the research questions and the categorization scheme used to extract the data. To mitigate these threats, the research questions and the categorization schemes were discussed among the authors.

\noindent{Conclusion validity:}
Concerning the subjectivity of the assessment of the PS's, the primary studies were reviewed by at least two authors to mitigate bias in data extraction. In case of disagreements, the researchers discussed these cases to reach consensus.


\section{Conclusion}
\label{sec:Conclusion}

In this paper, we mapped and reviewed the body of knowledge on behavior preservation in software refactoring.  We systematically reviewed 28 papers and classified them. This research sets out to aggregate, summarize, and discuss the practical approaches that  ensure behavior-preserving refactoring transformations. Our main findings show that (1) code artifacts have the main focus in refactoring literature, (2) some refactoring types were studied more frequently than others, (3) several behavior preservation approaches proposed in the literature including the concepts and techniques that guarantee program correctness when dealing with refactoring activities, the automated analyses that are proposed, and the manual analysis approach, and (4) the majority of the PSs empirically evaluate their approaches. This existing research evaluates the correctness of the transformation and whether or not these approaches lead to a safe and trustworthy refactoring.

\noindent{\textit{Lesson learned.} Research around behavior preservation of software refactoring has mainly focused on precondition-based strategy. However, other techniques such as graph-based have potential and might be more effective for particular problems. Consequently, current and future research in this area should explore the suitability of each technique based on the context and the possibility of incorporating several strategies to ensure the correctness of program transformation. Further, current refactoring engines are limited to certain features. Future research should strive to implement a full-featured refactoring engine to increase developers' trust in refactoring tools.}

\begin{appendices}
\section{Primary Studies}
\label{sec:Appendix}
List of accepted literature publications: 
{\footnotesize
\begin{enumerate}
\item[(PS1)] \bibentry{roberts1997refactoring}
\item[(PS2)] \bibentry{Mens2003FormalisingRW}
\item[(PS3)] \bibentry{Tip:2003:RGU:949305.949308}
\item[(PS4)] \bibentry{4026866}
\item[(PS5)] \bibentry{VanDerStraeten2007}
\item[(PS6)] \bibentry{Massoni:2008:FMP:1792838.1792873}
\item[(PS7)] \bibentry{article}
\item[(PS8)]  \bibentry{ubayashi2008contract}
\item[(PS9)] \bibentry{schafer2008sound}
\item[(PS10)] \bibentry{soares2009generating}
\item[(PS11)] \bibentry{4752842}  
\item[(PS12)] \bibentry{schafer2010specifying}
\item[(PS13)] \bibentry{5440166}    
\item[(PS14)]  \bibentry{tsantalis2010identification}
\item[(PS15)]  \bibentry{tip2011refactoring}
\item[(PS16)] \bibentry{soares2011making}
\item[(PS17)] \bibentry{6100067}
\item[(PS18)] \bibentry{6080784}
\item[(PS19)] \bibentry{SOARES20131006}
\item[(PS20)]  \bibentry{de2012language}
\item[(PS21)]  \bibentry{noguera2012refactoring}
\item[(PS22)]  \bibentry{thies2012refaflex}
\item[(PS23)] \bibentry{Mongiovi:2014:MRS:2664678.2664813}
\item[(PS24)]  \bibentry{najafi2016set}
\item[(PS25)] \bibentry{Horpcsi2017TrustworthyRV}
\item[(PS26)] \bibentry{7898404}
\item[(PS27)]  \bibentry{chen2018improving}
\item[(PS28)]  \bibentry{insa2018behaviour}

\end{enumerate}}
\end{appendices}


{\footnotesize\bibliography{mybibfile}}

\end{document}